# Tuning Nanocrystal Surface Depletion by Controlling Dopant Distribution as a Route Toward Enhanced Film Conductivity

Corey M. Staller[†], Zachary L. Robinson[∥], Ankit Agrawal[†], Stephen L. Gibbs[†], Benjamin L. Greenberg[‡], Sebastien D. Lounis[⌒,§], Uwe R. Kortshagen[‡], Delia J. Milliron[*,†]

[†]McKetta Department of Chemical Engineering, University of Texas at Austin, Austin, Texas, 78712-1589, United States

[∥] Department of Physics, University of Minnesota, Minneapolis, Minnesota, 55455, United States

[‡]Department of Mechanical Engineering, University of Minnesota, Minneapolis, Minnesota, 55455, United States

[⌒]The Molecular Foundry, Lawrence Berkeley National Laboratory, 1 Cyclotron Road, Berkeley, California 94720, United States

[§]Graduate Group in Applied Science & Technology, University of California, Berkeley, Berkeley, California, 94720, United States

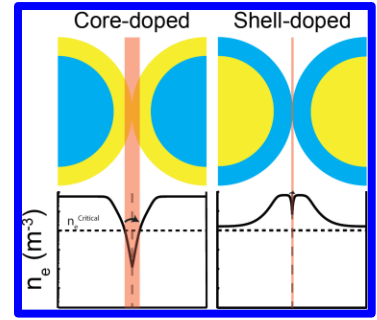

ABSTRACT: Electron conduction through bare metal oxide nanocrystal (NC) films is hindered by surface depletion regions resulting from the presence of surface states. We control the radial dopant distribution in tin-doped indium oxide (ITO) NCs as a means to manipulate the NC depletion width. We find in films of ITO NCs of equal overall dopant concentration that those with dopant-enriched surfaces show decreased depletion width and increased conductivity. Variable temperature conductivity data shows electron localization length increases and associated depletion width decreases monotonically with increased density of dopants near the NC surface. We calculate band profiles for NCs of differing radial dopant distributions and, in agreement with variable temperature conductivity fits, find NCs with dopant-enriched surfaces have narrower depletion widths and longer localization lengths than those with dopant-enriched cores. Following amelioration of NC surface depletion by atomic layer deposition of alumina, all films of equal overall dopant concentration have similar conductivity. Variable temperature conductivity measurements on alumina-capped films indicate all films behave as granular metals. Herein, we conclude that dopant-enriched surfaces decrease the near-surface depletion region, which directly increases the electron localization length and conductivity of NC films.

Transparent conductive oxide (TCO) thin films are of fundamental importance in the modern world due to their vast application in optoelectronic devices such as displays, solar cells, and electrochromic windows.[1,2] These applications require high conductivity, which has traditionally been achieved through vacuum deposition of amorphous or crystalline doped metal oxide films.[2] An effort to reduce manufacturing cost by moving away from vacuum deposition has motivated research on using films of colloidal nanocrystals (NCs) as TCO films. However, colloidal NCs are synthesized with long chain organic capping ligands resulting in spatial separations between neighboring NCs that act as tunneling barriers.[3–5] Carrier conduction in these systems occurs through a hopping mechanism, which is described by the Miller-Abraham model

$$\sigma \propto A exp\left(-\frac{2r_{ij}}{a}\right) exp\left(-\frac{E_{ij}}{k_B T}\right)$$

Where $\sigma$ is conductivity, $A$ is a material-dependent constant, $r_{ij}$ is the distance between sites $i$ and $j$, $a$ is the inverse of the wavefunction decay rate (called the electron localization length), $E_{ij}$ is the energetic barrier encountered moving from site $i$ to $j$, $k_B$ is the Boltzmann constant, and $T$ is temperature.[6] Inspection of the Miller-Abraham model leads to two obvious routes to improve NC film conductivity – reducing the distance between sites and lowering the energetic cost of hopping.[7,8]

Many efforts have focused on reducing the distance between NCs to improve electron transport through NC films. Initial work on colloidally synthesized NC films concentrated on exchanging the organic capping ligands used in synthesis with a variety of organic and inorganic ligands of different lengths or bonding arrangements to modify inter-NC charge transfer or hopping.[4,5,9,10] However, NC films with any ligands still in place are often too resistive to be suitable for device applications. Various ligand stripping and decomposition reactions have been developed to improve the conductivity of NC films.[5,11] Removal of ligands from NC surfaces often leads to several orders of magnitude increase in conductivity. While this strategy significantly improves film conductivity, removal of ligands exposes NC surfaces to adventitious chemical species, such as water that leads to hydroxylation, forming surface states that are difficult to control and can be harmful to charge transport through NC films.

One method to improve conduction through NC films is to use atomic layer deposition (ALD) to cap bare NC arrays with metal oxides, such as alumina ($Al_2O_3$).[12–16] Specifically, Thimsen et al. found that ZnO NC films with alumina capping layers had conductivity eight orders of magnitude higher than that of bare films.[13] This approach was later elaborated upon by Ephraim et al., who explained that the significant film performance improvement following alumina deposition is due to the removal of adsorbed water species by trimethylaluminum, the precursor used during alumina ALD.[14] These studies suggest that adsorbed water species play a direct role in affecting film conductivity but the mechanism by which adsorbed water species actually lead to reduced conductivity remains unexplored.[13,17] Recently, Zandi and Agrawal et. al. reported that electrochemical modulation of optical absorption in tin-doped indium oxide (ITO) NC films can be explained by the formation of a depletion region near NC surfaces.[17] We can therefore hypothesize that the enhanced electron transport in films whose surface hydroxyls have been eliminated is a result of alleviating depletion effects that were present due to the hydroxyl-associated surface states. Here we examine the role that depletion plays in inhibiting charge transport, and we study how the properties of NCs and their surfaces can be tuned to reduce depletion effects and improve transport.

Specifically, we report the influence of the intra-NC dopant distribution on conductivity of ITO NC films. Films comprised of NCs of similar size and overall tin concentration show a marked difference in film conductivity when the radial dopant distribution is manipulated. Bare NC films of a given overall dopant concentration exhibit higher conductivity, larger electron localization length, and lower contact resistance when dopant concentration is high near the NC surface. The dependence of electronic properties on dopant distribution is understood by examining how the intra-NC band profile is altered by dopant segregation in the presence of a depletion region near the NC surface. Following alumina ALD, films display comparable conductivity and contact resistance, independent of NC dopant profile, confirming that depletion-related resistance plays a dominant role in differentiating the electronic behavior of bare NC films of differing intra-NC dopant profiles.

**Experimental Procedures**
ITO NCs were synthesized using a two-step method adapted from the slow growth methods developed by Jansons et. al.[18] The dopant distribution was controlled by synthesizing ITO NC cores of a desired dopant concentration, which then undergo a washing procedure before reintroduction to a reaction flask for shell growth of desired dopant concentration and shell thickness (see SI Text 1 and Figure S2 for further synthetic details and X-ray diffraction (XRD)). This synthetic method leads to highly controlled core and shell sizes and low size polydispersity.

Core and overall particle sizes were measured by Scherrer analysis of the ITO (222) XRD peak and validated by scanning transmission electron microscopy (Figures S1-3). Dopant incorporation was quantified by elemental analysis using inductively coupled plasma-atomic emission spectroscopy (ICP-AES) for overall Sn dopant concentration and X-ray photoelectron spectroscopy (XPS) with an Al $K_\alpha$ source (1486.7 eV) to assess the near-surface Sn dopant concentration (Figure S4). Al $K_\alpha$ source energy corresponds to a photoelectron escape depth of about 1.5 nm.[19]

Colloidal NCs were spin-coated from a concentrated dispersion in a mixed solvent of hexane and octane (1:1) onto silicon and quartz substrates, yielding approximately 100 nm thick films. To enhance electron transport, the organic ligands used in colloidal NC synthesis were removed by an *in situ* ligand displacement with formic acid followed by a 60 minute anneal at 300°C in flowing nitrogen gas to decompose and desorb the remaining organic matter.[11] The resulting films were highly transparent at visible wavelengths (Figure S5). Scanning electron microscopy (SEM) images show densely packed films with direct contact between NCs and minimal cracking (Figure 1). Porosity of NC films prepared on silicon substrates was determined using ellipsometric porosimetry (EP) with toluene as the dielectric contrast solvent. EP data from 400 nm to 1000 nm wavelength was fit using software provided by JA Woollam and yielded consistent volume fractions between 0.72 and 0.78 for all films (Figure S6).

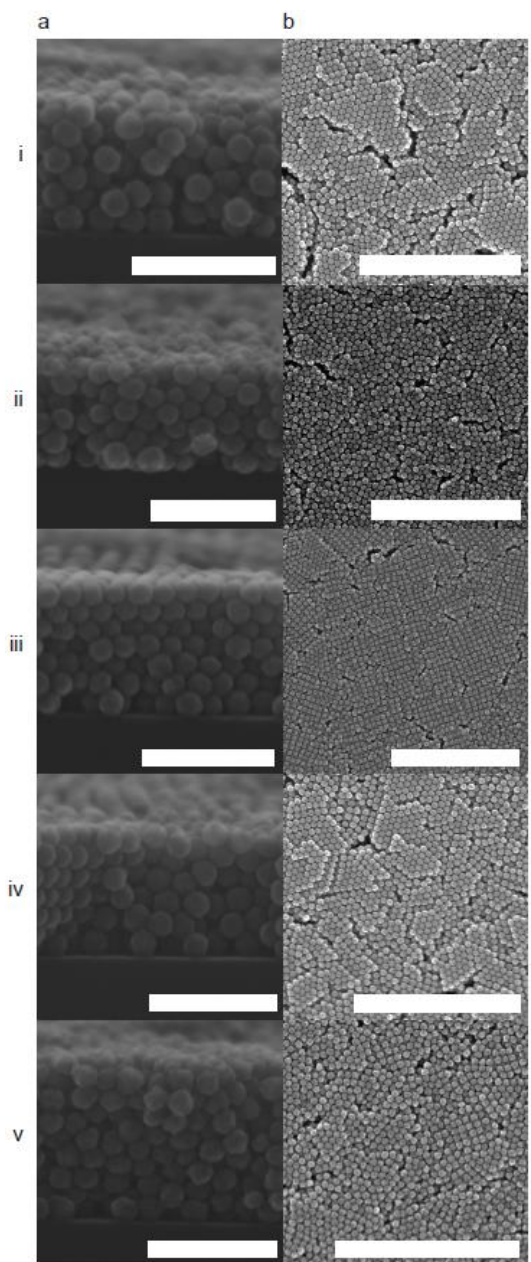

**Figure 1. Bare NC film on silicon SEM** cross-section (a) and top-down (b) for Core8 (i), Core5 (ii), Uniform (iii), Shell5 (iv), and Shell8 (v). Scale bars represent 100nm for cross-section and 500nm for top-down.

To understand the influence of dopant distribution on film electronic properties when surface depletion is suppressed, NC surfaces were passivated using 40 ALD cycles of alumina deposition. Deposition was carried out in a Savannah ALD chamber using previously reported methods.[14] Trimethylaluminum was used as the aluminum precursor and deposition was carried out at 180°C. These conditions correspond to a growth rate of about 0.11 nm per ALD cycle.[13] SEM showed 40 cycles of alumina deposition resulted in nearly complete infilling of NC films and deposition of a thin overlayer on the films (Figure S7). A similar approach was used by Ephraim et. al., who reported that deposition of alumina on surface segregated ITO NC films by ALD removes adsorbed water species and yields conductive ITO-alumina composites.[14]

To minimize aberrations in data, all analyses were conducted on samples that were exposed to ambient lab air for at least 23 days (Figure S8). Room temperature conductivity measurements were collected on an Ecopia Hall Effect measurement system (HMS-5000) in the 4-point probe Van der Pauw geometry. Gold spring-clip contacts were placed directly on the films and edge effects were minimized by isolating a uniform square region in the center of the film using a diamond scribe. Variable temperature conductivity measurements were conducted in a Physical Property Measurement System (PPMS) from as low as 2 K up to 300 K in both decreasing and increasing temperature directions. Ohmic contact was established using indium solder pads.

**Bare NC Films**

To investigate the role that dopant distribution plays in conductivity of ITO films, it was necessary to synthesize a series of NCs with similar size and overall dopant concentration (two properties that are known to affect conductivity)[20–22] but with variations in the radial profile of dopants. More specifically, thanks to unprecedented size and dopant incorporation control afforded by the synthetic methods advanced by Janson et al,[18] we varied the density of dopants radially within each NC while keeping the overall dopant concentration and NC diameter nearly constant at 3 at% and 20 nm, respectively. The five samples investigated here are uniformly-doped (Uniform), core-doped with an undoped shell: 5 at% core (Core5) and 8 at% core (Core8), and shell-doped with an undoped core: 5 at% shell (Shell5) and 8 at% shell (Shell8). NC sizes and dopant profile are summarized in Table 1. Discrepancies between nominal shell dopant concentration and that measured by XPS may be due to moderate redistribution of Sn. Despite this, comparing tin content by XPS and ICP-AES shows significant dopant segregation for all core-shell samples and a clear trend of increasing near-surface dopant concentration from Core8 to Shell8.

**Table 1. ITO NC core-shell structure.** Core and overall NC sizes by Scherrer analysis, overall tin dopant concentration by ICP-AES and near surface tin dopant concentration by XPS.

| Sample | Dopant Distribution | Core Diameter (nm) | Overall Diameter (nm) | at% Sn by ICP-AES | at% Sn by XPS |
|---|---|---|---|---|---|
| Core8 | Core-doped | 14.4 | 20.8 | 3.3±0.2 | 2.2 |
| Core5 | Core-doped | 16.5 | 20.8 | 3.1±0.3 | 2.4 |
| Uniform | Uniform | 16.6 | 20.2 | 3.0±0.1 | 3.9 |
| Shell5 | Surface-doped | 15.1 | 20.7 | 3.1±0.2 | 6.0 |
| Shell8 | Surface-doped | 16.8 | 19.8 | 2.5±0.1 | 6.3 |

We measure conductivity of bare films following prolonged ambient lab air exposure to minimize variations in surface chemistry from sample to sample. Bare film conductivity is shown in Figure 2. Conductivity of all samples of equal overall dopant concentration shows an exponential dependence on the near-surface dopant concentration. Core8 and Core5 exhibit the lowest average conductivity of the series at 0.154 S-cm$^{-1}$ and 0.168 S-cm$^{-1}$, respectively. As dopants are placed closer to the surface, the conductivity more than doubles upon reaching uniform distribution, where the measured conductivity was 0.343 S-cm$^{-1}$. Finally, Shell5 shows the highest conductivity of the series at 0.901 S-cm$^{-1}$, representing a nearly nine-time increase from the lowest conductivity sample, Core8. We note that while Shell8 has the highest dopant concentration on the surface, it shows a significantly lower conductivity than expected based on the trend observed for the four other samples. One possible explanation for this deviation may be Shell8 having the a significantly lower overall dopant concentration, however the material-dependent constant, $A$, in the Miller-Abraham model and its dependence on dopant concentration are highly uncertain.[22–25] We refrain from analyzing the room temperature conductivity of bare films of Shell8 for this reason.

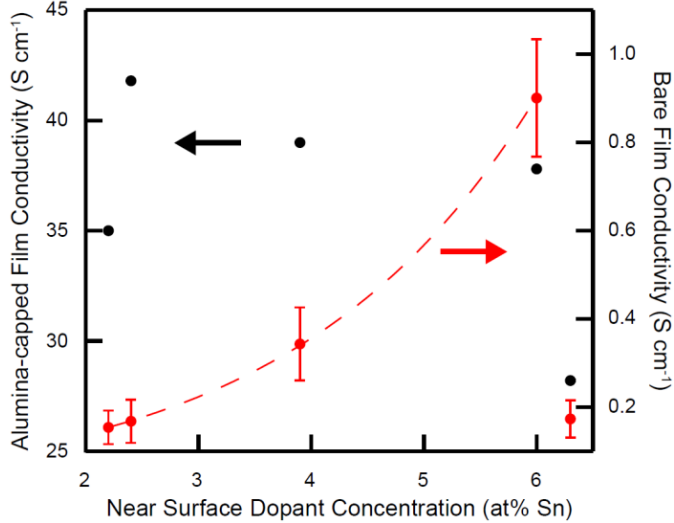

**Figure 2. Film conductivity.** Room temperature conductivity for bare ITO NC films of equal overall dopant concentration shows a strong exponential dependence on the near surface dopant concentration, as measured by XPS. This dependence is not observed in the room temperature conductivity of alumina-capped ITO NC films. Dashed line shows an exponential fit of bare film conductivity.

Comparing room temperature conductivity of films is useful in determining the optimal material for a device, but gives little insight into the differences in electron transport physics underlying these differences. Analysis of the underlying physics requires films to be viewed as a random resistor network composed of randomly positioned bonds, i.e. conduction pathways, each with a finite resistance, $R_{bond}$. For NC films with much lower conductivity than their bulk analogue, bond resistance is approximately equal to the contact resistance, $R_C$, which describes the tunneling resistance between NCs,[16] and is calculated from the links and nodes model in three dimensions as

$$R_C \approx R_{bond} = \frac{(\varphi - \varphi_0)^{1.9}}{2\sigma r_0}$$

where $\sigma$ is the film conductivity, $\varphi$ is the NC volume fraction, $\varphi_0$ is the percolation threshold, and $r_0$ is the NC radius.[26,27] We assume the percolation threshold to be that of randomly packed spheres, approximately 0.2.[28] Inter-NC contact resistances are reported in Table 2. When $R_C$ is greater than the critical tunneling resistance, a material behaves as an insulator and conduction is dominated by a hopping mechanism.[29] $R_C$ is well above the critical tunneling resistance for all NC films indicating that electrons must hop between NCs (SI Text 2).

**Table 2. Localization length and contact resistance.** Bare films: localization length determined by carrier concentration profile simulations, localization length determined by ES-VRH-GD fits, and contact resistance found using the links and nodes model. Alumina-capped films: localization length determined by carrier concentration profile simulations and metallic grain size and contact resistance found by granular metal fits.

|  | Bare | | | Alumina-capped | | |
| --- | --- | --- | --- | --- | --- | --- |
| Sample | Simulated Localization Length (nm) | Localization Length (nm) | Contact Resistance (kΩ) | Simulated Localization Length (nm) | Metallic Grain Size (nm) | Contact Resistance (kΩ) |
| Core8 | 18.2 | 17.9 | 1240 | ≥ 20.8 | 24.8 | 20.2 |
| Core5 | 19.2 | 17.8 | 915 | ≥ 20.8 | 22.0 | 19.2 |
| Uniform | 19.8 | 18.2 | 504 | ≥ 20.2 | 22.1 | 20.8 |
| Shell5 | 20.6 | 21.5 | 152 | ≥ 20.7 | 22.7 | 19.8 |
| Shell8 | ≥ 19.8 | 24.7 | 974 | ≥ 19.8 | 24.8 | 24.0 |

In films with no necking between NCs, $R_C$ is primarily defined by a tunneling junction with resistance proportional to $\exp(s\sqrt{2m^*U_0}/\hbar)$ where $s$ is the barrier width, $U_0$ is barrier height, $m^*$ is the effective mass of an electron (0.4$m_e$ for ITO), and $\hbar$ is Planck's constant.[22] We use the magnitude of $R_C$ as a metric to estimate the tunneling width. The barrier height, equal to the work function at the NC surface, is determined by the surface state energy due to Fermi level pinning and is assumed to be equal for all samples. One should note that this analysis is a simple estimate as the tunneling resistance has a pre-exponential factor that may have a dependence on overall dopant concentration or dopant distribution. However, it is clear from the order of magnitude difference in $R_C$ between Shell5 and Core8 that NC dopant distribution significantly affects the tunneling barrier.

We performed variable temperature conductivity measurements on the bare NC films to gain further insight on how intra-NC dopant profiles influence the inter-NC tunneling junctions (Figure 3a). Conductivity increases monotonically with increasing temperature for all films, characteristic of electrons conducting through a hopping mechanism. The temperature dependence of conductivity in an electron hopping regime is described by

$$\sigma(T) = \sigma_0 * \exp\left(-\left(\frac{T_0}{T}\right)^m\right)$$

Where $\sigma_0$ is treated as a material-dependent constant, $T_0$ is a characteristic temperature, and $m$ depends on the specific hopping mechanism.[24] Zabrodskii analysis indicates $m$ values of nearly 0.78 for all bare samples we measured (Figure S9).[30] Atypical values of $m$ (those other than 0.25, 0.5, or 1) were investigated by Houtepen et. al. and were explained by temporal broadening of energy levels within the density-of-states.[31] This broadening depends on the temperature dependence of heat capacity for the active material. While Houtepen et. al. assumed a constant heat capacity and found an m-value of 0.66, we used a power law fit to the ITO heat capacity ($C_p \propto CT^p$) to capture this temperature dependence in the model. Using this power-law heat capacity fit we derived the Efros-Shklovskii variable-range hopping with a Gaussian dispersion of energy levels (ES-VRH-GD) to have an $m$ value of 0.78 (Figure S10, SI Text 3). The ES-VRH-GD characteristic temperature in this regime is then defined as

$$T_0 = \left(\frac{3.15e^4}{4\pi^2\varepsilon^2 k_B C a^2}\right)^{\frac{1}{3}}$$

Where $C \approx 27$ is the heat capacity power-law coefficient for ITO, $e$ is the electron charge, $\varepsilon$ is the dielectric constant of the film, and $a$ is the electron localization length. The film effective dielectric constant was calculated using methods developed by Reich and Shklovskii.[32] Here, the electron localization length defines the diameter of a sphere within which mobile electrons are confined at 0K. Fits to variable temperature data and the corresponding localization lengths are reported in Figure 3b and Table 2, respectively.

ES-VRH-GD fits to variable temperature conductivity data indicate monotonic increase of the electron localization length as dopants move toward the surface. Core8 NC films exhibit a localization length nearly 3 nm smaller than the NC diameter while Shell8 shows slight delocalization of electrons beyond the size of the NCs. The monotonic growth of localization length with increasing overall dopant concentration has been established as a signature of approaching the metal-insulator-transition in NC films.[16,22,26] Interestingly, despite having the lowest overall dopant concentration, Shell8 shows the largest localization length of the samples measured here. Previously, studies of the connection between localization and dopant concentration considered only films of uniformly doped NCs and those with passivated surfaces. Here we reveal that the more relevant property for bare NC films is the concentration of dopants in the near surface region. This is because greater (lesser) dopant density at the surface reduces (increases) the effects of surface depletion. Thus, films of low overall dopant concentration NCs can be engineered to produce a large localization length by controlling the dopant distribution.

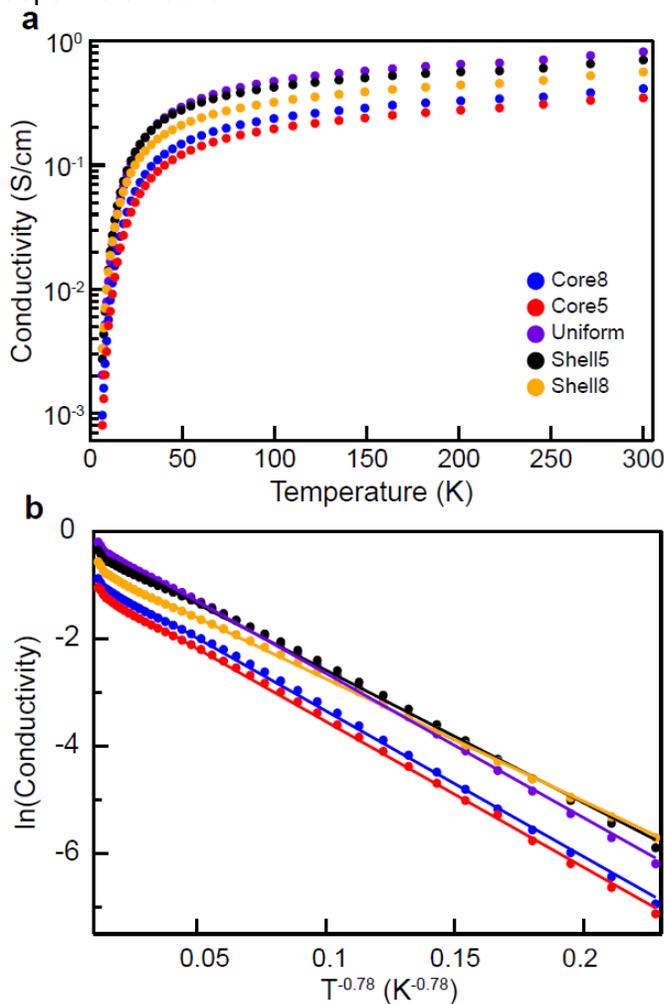

**Figure 3. Temperature dependence of electron transport in bare ITO NC films.** Conductivity v. Temperature (a) and Efros-Shklovskii variable range hopping with a Gaussian dispersion of energy levels fit for bare ITO NC films (b). Markers indicate experimental data and lines show fits to ES-VRH-GD.

To further understand the trend of localization length with changing dopant distribution, we simulated the band profiles within isolated ITO NCs with radially controlled dopant distribution in the presence of surface states that are approximated to be 0.2 eV below the conduction band minimum of indium oxide (Figure 4, SI Text 4 and Figure S11).[17] The inequality of electrochemical potential on the surface and in the bulk drives electrons from the NC into the unoccupied surface states, pinning the Fermi energy at the surface state energy. This occurs in any semiconductor NC when the surface state lies below the bulk Fermi energy. In uniformly-doped ITO NCs (Figure 4ci), the band profile is easily understood as a radial depletion region near the NC surface. When dopants are segregated, the band profile becomes significantly more complex. Tin dopants decrease the electronic band gap of indium oxide while also increasing the optical band gap due to state filling, i.e. the Burstein-Moss effect.[33,34] This means that in addition to band bending at the surface, band bending will occur near the interface of doped and undoped regions within the NC. In core-doped ITO NCs (Figure 4a,bi), a relatively low density of charged defects near the surface results in a wide depletion region that extends to the doped core. In contrast, surface-doped ITO NCs (Figure 4d,ei) have a high density of charged defects near the surface, resulting in a sharper, narrower depletion region, which we have correlated with an expanded localization length in films fabricated from these NCs.

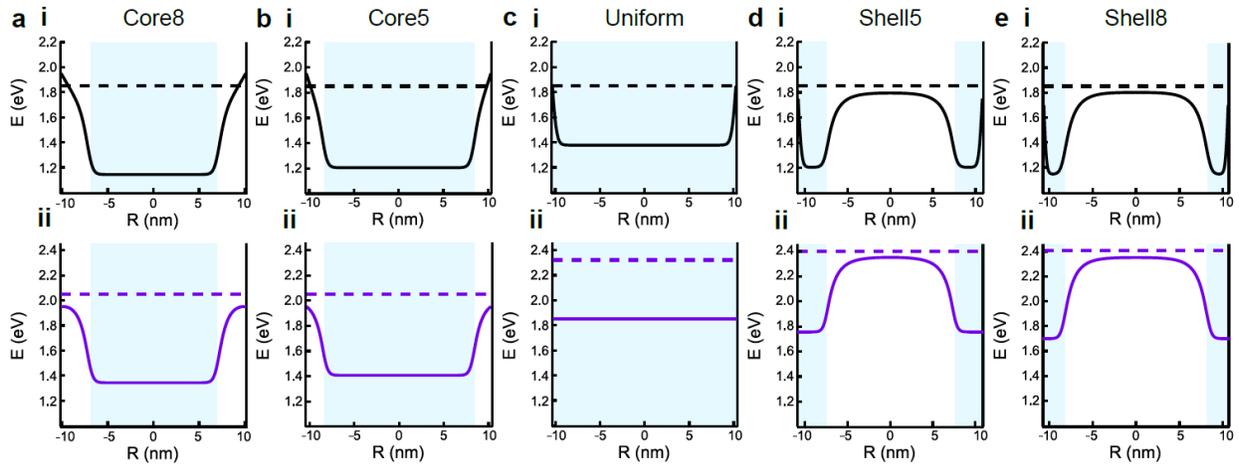

**Figure 4. Simulated band profiles.** Intra-NC band profile for Core8 (a), Core5 (b), Uniform (c), Shell5 (d), and Shell8 (e) with a surface potential 0.2 eV below the flat band potential of indium oxide (a-e i) and equal to the flat band potential of the shell species (a-e ii). The latter represents the absence of surface-state induced depletion. In all cases, the horizontal dashed line is the Fermi level and the blue shaded region indicates the region of enriched dopants. R = 0 denotes the center of a NC and maximum |R| denotes the surface.

We simulated electron concentration profiles to visualize how these band profiles influence the electron localization length for each dopant distribution in a NC-NC tunneling junction (Figure 5 a-e). The electron localization length is defined here as the diameter of a sphere containing all space with an electron concentration of greater than $10^{25}$ m$^{-3}$, the critical electron concentration for metallic behavior in ITO according to the Mott criterion.[35] Using this definition, we examine the electron localization compression due to depletion, which is simply the difference between the physical diameter of the NC and the electron localization length. As shown in Figure 5 a,b, the extended surface depletion region in core-doped samples leads to significant localization length shrinkage in Core8 and Core5 of 2.6 nm and 1.6 nm, respectively. For uniform dopant distribution (Figure 5 c), the localization length is compressed only slightly, about 0.4 nm. Finally, when the majority of dopants are near the surface (Figure 5 d,e), the localization volume is approximately the size of the NC with Shell5 having 0.1 nm compression and Shell8 being fully delocalized within the NC. The definition of electron localization length in our simulations is simplified by ignoring wavefunction decay beyond the metallic region of the NC. Despite this, the trend observed in simulations is mirrored in localization lengths determined by analysis of variable temperature conductivity data (Table 2). Without considering wavefunction decay beyond the physical NC dimensions our simulation cannot describe electron localization lengths greater than the NC diameter; however, the simulated intra-NC electron concentration profile for Shell8 agrees with the possibility for electron delocalization outside of the NC, as suggested by the experimental variable temperature conductivity data.

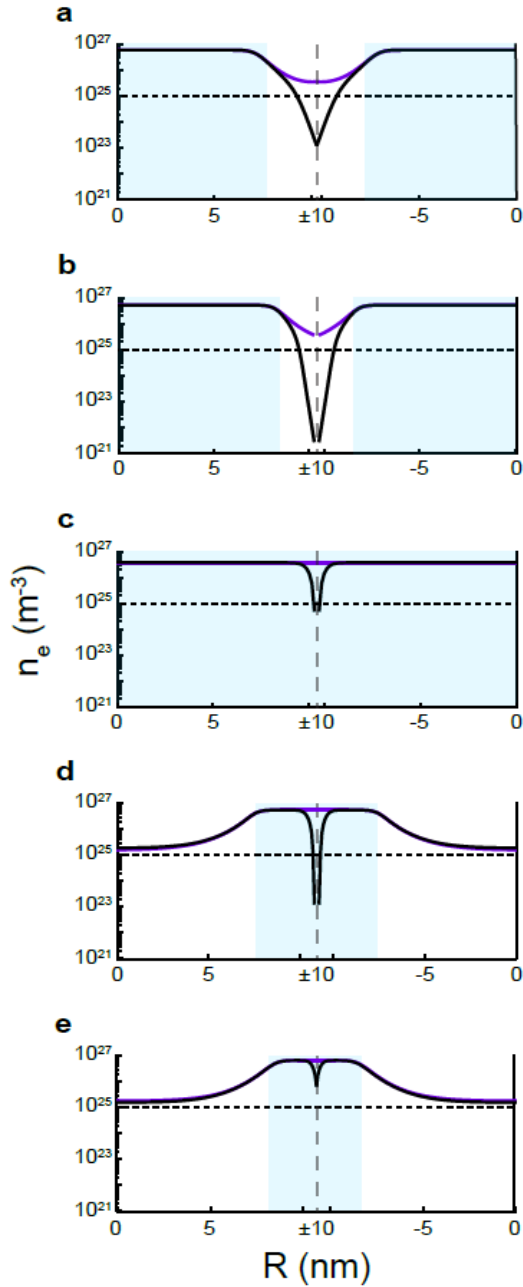

**Figure 5. Simulated electron concentration profiles.** Radial electron concentration profile for Core8 (a), Core5 (b), Uniform (c), Shell5 (d), and Shell8 (e) with a surface potential 0.2 eV below the flat band potential of indium oxide (Black) and equal to the flat band potential of the shell species (Purple). Vertical dashed line is where neighboring NCs touch and horizontal dashed line is the critical carrier concentration. In all cases, the blue shaded region indicates the region of enriched dopants. R = 0 denotes the center of a NC and maximum |R| denotes the surface.

The importance of the localization length and its magnitude relative to the NC size is apparent as we return to analyze the limiting factors in achieving high conductivity NC films. The contact resistance, analyzed as a tunneling junction, is $R_C \propto \exp(s\sqrt{2m^*U_0}/\hbar)$. Again, $s$ defines the separation between the edges of electron localization volumes. Using a simple model of two NCs in contact at a point, and electronically connected by a tunnel junction due to depletion, s is simply the electron localization compression discussed above (Figure 5). From our analysis, as the dopant profile moves toward the surface, there is monotonic decrease in s. This diminished barrier width implies a decrease in contact resistance, in agreement with experimental results up to Shell8.

**Alumina-capped Films**
Following alumina deposition, the conductivity of all films increased significantly as shown in Figure 2. Capped-NC films of equal dopant concentration exhibit nearly identical conductivity while, as in the bare NC film case, Shell8 shows the lowest conductivity. To explain the anomalous conductivity exhibited by Shell8 we return to the random resistor network picture, where the film conductivity is inversely proportional to the contact resistance between neighboring NCs.[16,26,27] In the absence of depletion, neighboring NCs are treated simply as doped semiconductor spheres in contact where the resistance between two NCs is inversely proportional to the square of the Fermi wavevector.[22] Due to the lower overall dopant concentration in Shell8, the Fermi wavevector is expected to be lower resulting in lower film conductivity.

We performed variable temperature conductivity measurements on NC films after alumina capping to gain insight on electron transport barriers (Figure 6a). While conductivity of alumina-capped films increases monotonically with increasing temperature for all films, conductivity is consistent with a logarithmic dependence on temperature. Conduction of this type has been observed in granular ITO thin films and is ascribed to a granular metal conduction mechanism.[36–38] We note that while hopping conduction is a description of electrons overcoming a Coulombic blockage by undergoing hops of temporally varying distances, the granular metal conduction model describes the competition between Coulombic barriers and NC-NC coupling. Granular metal conduction depends on the grain (NC) size and the degree of coupling, reflected in tunneling conductance between neighboring NCs, rather than the electron localization length. The temperature dependence of conductivity in granular metals is described by

$$\sigma = \sigma_0 \left(1 + \frac{1}{2\pi g_T d} \ln\left(\frac{k_B T}{g_T E_C}\right)\right)$$

where $\sigma_0 = g_T \left(\frac{2e^2}{\hbar}\right)\alpha^{2-d}$, $g_T$ is the non-dimensional tunneling conductance between grains, $\alpha$ is the grain diameter, $d$ is the system dimensionality, and $E_C = \frac{e^2}{2\pi\varepsilon\alpha}$ is the charging energy of a grain.[29,36] Metallic grain size and tunneling conductance were calculated from the slope and intercept, respectively, of $\sigma$ vs $\ln(T)$ as shown in Figure 6b. Metallic grain size and tunneling conductance are reported in Table 2. This model is valid for all films measured as the tunneling conductance is below the intra-NC conductance and above the critical tunneling conductance, the inverse of the previously discussed critical tunneling resistance, and all fitted temperatures are above the quantum limit (SI Text 5).

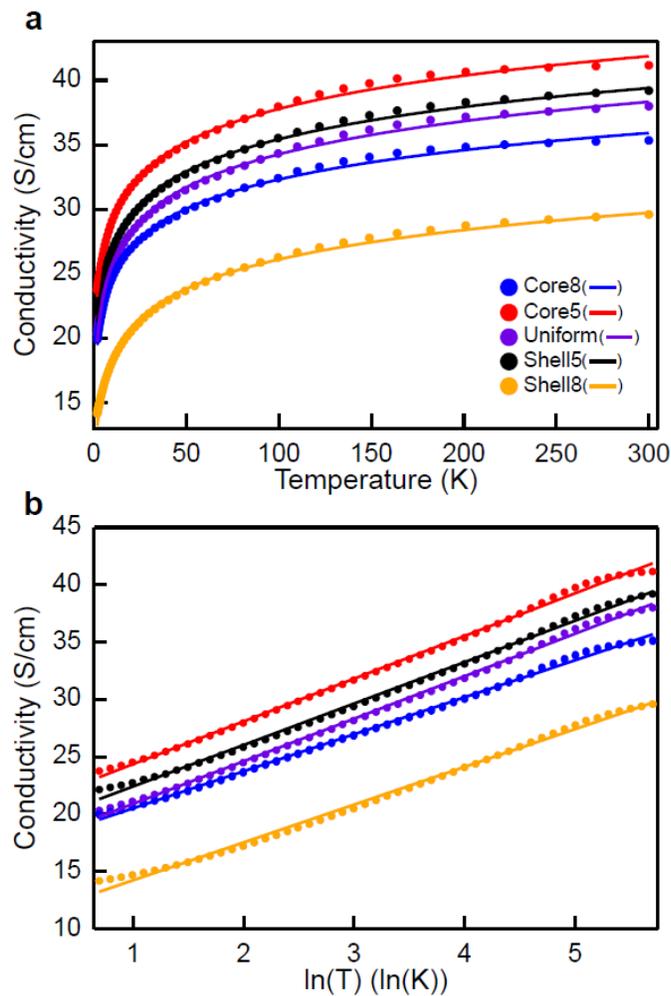

**Figure 6. Temperature dependence of electron transport in alumina-capped ITO NC films.** Conductivity v. Temperature (a) and granular metal conduction mechanism for alumina-capped ITO NC films (b). Markers indicate experimental data and lines show fits to the granular metal conduction mechanism.

The grain size in ITO NC films with alumina capping does not show a clear dependence on dopant distribution as localization length did in bare films, but rather becomes approximately equal to the NC diameter for all samples based on granular metal fits of variable temperature conductivity data. Simulated band profiles for passivated surface ITO NCs (Figure 4a-eii) show the electron concentration exceeds the critical value throughout the NCs despite the presence of the core-shell interface band bending. Shell- and uniformly-doped NCs (Figure 4c,d,eii) show a flat band on the surface while in core-doped NCs the accumulation region from the core has not yet reached flat band at the NC surface (Figure 4a,bii). Examination of electron concentration profiles (Figure 5) provides an explanation for the experimental observation that grain size does not show a clear dependence of dopant distribution. As shown in the figure, the localization length expands upon passivation of the surface, eliminating the separation between electron localization volumes (Figure 5). NC localization volumes are now metallic spheres touching via point contacts, which have a significantly lower contact resistance due to a decreased tunneling barrier width. Dopant profile dependence is observed in neither grain size nor contact resistance following surface passivation by ALD.

In conclusion, dopant distribution within NCs was shown to be an effective means to influence the electronic properties of bare ITO NC films (including overall conductivity, contact resistance, and electron localization length) at a constant overall tin concentration. These effects were understood based largely on the variations dopant distribution effected on the near-surface depletion layer. The model typically used to analyze variable temperature conductivity assumes full dopant ionization and uniform electron distribution within NCs, which is obviously not a fully physical description of our materials. However, no established theories describing electron conduction through NC films explicitly incorporate the potential for dopant segregation or intra-NC band bending. The assumptions of uniform electron distribution and full dopant ionization create some uncertainty in the meaning of the ES-VRH-GD derived localization length. However, considering the localization lengths derived from simulated electron concentration profiles agree well with the values derived from these fits to the experimental data suggests that any error caused by these assumptions is small compared to the influence of dopant segregation on transport properties. Conclusions drawn here should be viewed as a general case for doped semiconductors in the presence of surface states as our simulations and their interpretations are applicable across a range of systems.

We have examined the influence of dopant distribution within ITO NCs on the conductivity of NC films while also considering the presence of surface defects and the potential to passivate surface defects using ALD. Intra-NC dopant distribution plays a strong role in determining macroscopically observable electronic properties, such as film conductivity, and microscopic electronic properties, such as localization length and contact resistance, of bare NC films. The influence of dopant distribution on the properties listed above is the result of modulating surface depletion as these effects are eliminated following the deposition of alumina. These experimental results were supported by simulations of intra-NC band profiles, which showed identical trends in electron localization and its implications on contact resistance. Dopant distribution engineering offers a promising route through surface modification to improve the conductivity of NC films for device applications. Using intra-NC dopant distribution to tune surface depletion additionally creates interesting new avenues of study regarding phenomena and applications that depend on the interaction between the conduction electrons and the surrounding environments such as electrochromic devices, plasmonics materials, sensors, and catalysts.

ASSOCIATED CONTENT
Details of nanocrystal synthesis, nanocrystal characterization (STEM and XRD), Film characterization (XPS, visible spectroscopy, EP, alumina-capped SEM, and air stability), Conduction mechanism details (critical tunneling resistance, Zabrodskii analysis, heat capacity fit, derivation of Efros-Shklovskii variable range hopping with Gaussian broadening of energy levels, and granular metal model validity), and details of band profile simulations.

AUTHOR INFORMATION
Corresponding Author
E-mail: milliron@che.utexas.edu
Notes:
The authors declare no competing financial interest.

ACKNOWLEDGMENTS
This research was supported by the National Science Foundation (NSF), including NASCENT, an NSF ERC (EEC-1160494, C.M.S.), CHE-1609656 (A.A.), the University of Minnesota MRSEC (DMR-1420013; Z.L.R., G.L.G., U.R.K.), a Graduate Research Fellowship under Award

Number (DGE-1610403, S.L.G.), and the Welch Foundation (F-1848). This work was performed in part at the Molecular Foundry, Lawrence Berkeley National Laboratory, which is supported by the Office of Science, Office of Basic Energy Sciences, of the U.S. Department of Energy (DOE) under Contract No. DE-AC02-05CH11231. S.D.L. was supported by a DOE Early Career Research Program grant to D.J.M.

REFERENCES

(1) Barquinha, P.; Martins, R.; Pereira, L.; Fortunato, E. *Transparent Oxide Electronics*; John Wiley & Sons, Ltd: Chichester, UK, 2012.
(2) Ellmer, K. Past Achievements and Future Challenges in the Development of Optically Transparent Electrodes. *Nat. Photonics* **2012**, *6*, nphoton.2012.282.
(3) Kim, J.-Y.; Kotov, N. A. Charge Transport Dilemma of Solution-Processed Nanomaterials. *Chem. Mater.* **2014**, *26*, 134–152.
(4) Pham, H. T.; Jeong, H.-D. Newly Observed Temperature and Surface Ligand Dependence of Electron Mobility in Indium Oxide Nanocrystals Solids. *ACS Appl. Mater. Interfaces* **2015**, *7*, 11660–11667.
(5) Zarghami, M. H.; Liu, Y.; Gibbs, M.; Gebremichael, E.; Webster, C.; Law, M. P-Type PbSe and PbS Quantum Dot Solids Prepared with Short-Chain Acids and Diacids. *ACS Nano* **2010**, *4*, 2475–2485.
(6) Miller, A.; Abrahams, E. Impurity Conduction at Low Concentrations. *Phys. Rev.* **1960**, *120*, 745–755.
(7) Talapin, D. V.; Lee, J.-S.; Kovalenko, M. V.; Shevchenko, E. V. Prospects of Colloidal Nanocrystals for Electronic and Optoelectronic Applications. *Chem. Rev.* **2010**, *110*, 389–458.
(8) Zabet-Khosousi, A.; Dhirani, A.-A. Charge Transport in Nanoparticle Assemblies. *Chem. Rev.* **2008**, *108*, 4072–4124.
(9) Talapin, D. V.; Murray, C. B. PbSe Nanocrystal Solids for N- and p-Channel Thin Film Field-Effect Transistors. *Science* **2005**, *310*, 86–89.
(10) Liu, W.; Lee, J.-S.; Talapin, D. V. III–V Nanocrystals Capped with Molecular Metal Chalcogenide Ligands: High Electron Mobility and Ambipolar Photoresponse. *J. Am. Chem. Soc.* **2013**, *135*, 1349–1357.
(11) Garcia, G.; Buonsanti, R.; Runnerstrom, E. L.; Mendelsberg, R. J.; Llordes, A.; Anders, A.; Richardson, T. J.; Milliron, D. J. Dynamically Modulating the Surface Plasmon Resonance of Doped Semiconductor Nanocrystals. *Nano Lett.* **2011**, *11*, 4415–4420.
(12) Liu, Y.; Tolentino, J.; Gibbs, M.; Ihly, R.; Perkins, C. L.; Liu, Y.; Crawford, N.; Hemminger, J. C.; Law, M. PbSe Quantum Dot Field-Effect Transistors with Air-Stable Electron Mobilities above 7 Cm2 V−1 S−1. *Nano Lett.* **2013**, *13*, 1578–1587.
(13) Thimsen, E.; Johnson, M.; Zhang, X.; Wagner, A. J.; Mkhoyan, K. A.; Kortshagen, U. R.; Aydil, E. S. High Electron Mobility in Thin Films Formed via Supersonic Impact Deposition of Nanocrystals Synthesized in Nonthermal Plasmas. *Nat. Commun.* **2014**, *5*.
(14) Ephraim, J.; Lanigan, D.; Staller, C.; Milliron, D. J.; Thimsen, E. Transparent Conductive Oxide Nanocrystals Coated with Insulators by Atomic Layer Deposition. *Chem. Mater.* **2016**, *28*, 5549–5553.
(15) Liu, Y.; Gibbs, M.; Perkins, C. L.; Tolentino, J.; Zarghami, M. H.; Bustamante, J.; Law, M. Robust, Functional Nanocrystal Solids by Infilling with Atomic Layer Deposition. *Nano Lett.* **2011**, *11*, 5349–5355.
(16) Greenberg, B. L.; Robinson, Z. L.; Reich, K. V.; Gorynski, C.; Voigt, B. N.; Francis, L. F.; Shklovskii, B. I.; Aydil, E. S.; Kortshagen, U. R. ZnO Nanocrystal Networks Near the Insulator–Metal Transition: Tuning Contact Radius and Electron Density with Intense Pulsed Light. *Nano Lett.* **2017**.

(17) Zandi, O.; Agrawal, A.; Shearer, A. B.; Gilbert, L. C.; Dahlman, C. J.; Staller, C. M.; Milliron, D. J. Impacts of Surface Depletion on the Plasmonic Properties of Doped Semiconductor Nanocrystals. *ArXiv170907136 Cond-Mat Physicsphysics* **2017**.
(18) Jansons, A. W.; Hutchison, J. E. Continuous Growth of Metal Oxide Nanocrystals: Enhanced Control of Nanocrystal Size and Radial Dopant Distribution. *ACS Nano* **2016**, *10*, 6942–6951.
(19) Hewitt, R. W.; Winograd, N. Oxidation of Polycrystalline Indium Studied by X-Ray Photoelectron Spectroscopy and Static Secondary Ion Mass Spectroscopy. *J Appl Phys* **1980**, *51*, 2620–2624.
(20) Kang, M. S.; Sahu, A.; Norris, D. J.; Frisbie, C. D. Size-Dependent Electrical Transport in CdSe Nanocrystal Thin Films. *Nano Lett.* **2010**, *10*, 3727–3732.
(21) Liu, Y.; Gibbs, M.; Puthussery, J.; Gaik, S.; Ihly, R.; Hillhouse, H. W.; Law, M. Dependence of Carrier Mobility on Nanocrystal Size and Ligand Length in PbSe Nanocrystal Solids. *Nano Lett.* **2010**, *10*, 1960–1969.
(22) Chen, T.; Reich, K. V.; Kramer, N. J.; Fu, H.; Kortshagen, U. R.; Shklovskii, B. I. Metal-Insulator Transition in Films of Doped Semiconductor Nanocrystals. *Nat. Mater.* **2015**, *advance online publication*.
(23) Arginskaya, N. V.; Kozub, V. I. Potential Influence of Pre-Exponential Factors on the Temperature Dependence of Variable-Range Hopping Conductivity. *Sov. J. Exp. Theor. Phys.* **1994**, *79*, 466–472.
(24) Shklovskii, B. I.; Efros, A. L. *Electronic Properties of Doped Semiconductors*; Springer-Verlag Berlin Heidelberg GmbH.
(25) Yildiz, A.; Serin, N.; Serin, T.; Kasap, M. Crossover from Nearest-Neighbor Hopping Conduction to Efros–Shklovskii Variable-Range Hopping Conduction in Hydrogenated Amorphous Silicon Films. *Jpn. J. Appl. Phys.* **2009**, *48*, 111203.
(26) Lanigan, D.; Thimsen, E. Contact Radius and the Insulator–Metal Transition in Films Comprised of Touching Semiconductor Nanocrystals. *ACS Nano* **2016**, *10*, 6744–6752.
(27) Shklovskiĭ, B. I.; Éfros, A. L. Percolation Theory and Conductivity of Strongly Inhomogeneous Media. *Sov. Phys. Uspekhi* **1975**, *18*, 845.
(28) Powell, M. J. Site Percolation in Randomly Packed Spheres. *Phys. Rev. B* **1979**, *20*, 4194–4198.
(29) Beloborodov, I. S.; Lopatin, A. V.; Vinokur, V. M. Universal Description of Granular Metals at Low Temperatures: Granular Fermi Liquid. *Phys. Rev. B* **2004**, *70*, 205120.
(30) Zabrodskii, A. G. The Coulomb Gap: The View of an Experimenter. *Philos. Mag. Part B* **2001**, *81*, 1131–1151.
(31) Houtepen, A. J.; Kockmann, D.; Vanmaekelbergh, D. Reappraisal of Variable-Range Hopping in Quantum-Dot Solids. *Nano Lett.* **2008**, *8*, 3516–3520.
(32) Reich, K. V.; Shklovskii, B. I. Dielectric Constant and Charging Energy in Array of Touching Nanocrystals. *Appl. Phys. Lett.* **2016**, *108*, 113104.
(33) Berggren, K.-F.; Sernelius, B. E. Band-Gap Narrowing in Heavily Doped Many-Valley Semiconductors. *Phys. Rev. B* **1981**, *24*, 1971–1986.
(34) Hamberg, I.; Granqvist, C. G.; Berggren, K.-F.; Sernelius, B. E.; Engström, L. Band-Gap Widening in Heavily Sn-Doped $In_2O_3$. *Phys. Rev. B* **1984**, *30*, 3240–3249.
(35) Mott, N. F. Conduction in Non-Crystalline Systems. *Philos. Mag.* **1968**, *17*, 1259–1268.
(36) Beloborodov, I. S.; Lopatin, A. V.; Vinokur, V. M.; Efetov, K. B. Granular Electronic Systems. *Rev. Mod. Phys.* **2007**, *79*, 469–518.
(37) Lin, J.-J.; Li, Z.-Q. Electronic Conduction Properties of Indium Tin Oxide: Single-Particle and Many-Body Transport. *J. Phys. Condens. Matter* **2014**, *26*.
(38) Zhang, Y.-J.; Li, Z.-Q.; Lin, J.-J. Logarithmic Temperature Dependence of Hall Transport in Granular Metals. *Phys. Rev. B* **2011**, *84*, 052202.

# Supplemental Information:
## SI Text 1. Methods.
### Synthesis of ITO NCs

ITO NCs were synthesized by modification of methods published by the Hutchison group.[1,2] NC cores were synthesized by adding 4.7 mmol of metal precursor (In(III)Acetate$_3$ and Sn(IV)Acetate$_4$) to 10 mL of oleic acid in a round bottom flask. This will be referred to as the precursor flask. The precursor flask is then put under vacuum and heated to 110°C for 1 hour with one pump/purge midway through the hour. The precursor flask is then put under nitrogen and heated to 150°C for 2 hours to generate In- and Sn-oleate. Concurrently, 12 mL of oleyl alcohol is put in a second round bottom flask, called the reaction flask. The reaction flask is put under vacuum and heated to 150°C for 2 hours with one pump/purge midway through. The reaction flask is then heated to 290°C under nitrogen. Once the In- and Sn-oleate reaction has finished, the contents of the precursor flask are pulled into a syringe for slow injection into the reaction flask. The injection rate is set to 0.2 mL/min and the injection volume depends on the desired core size. Following the injection the reaction flask is allowed to stay at 290°C for 20 min before being cooled to room temperature. NC cores are washed by 5 cycles of crashing NCs with ethanol, centrifuging at 7500 RPM, and redispersing in hexane. For shelling, the NC core dispersion in hexane is added to the reaction flask with oleyl alcohol and the synthesis is conducted in an identical manner as the core synthesis with volume of reaction mixture injected dictating shell thickness.

### Scanning electron microscopy (SEM) and scanning transmission electron microscopy (STEM) measurements

SEM and STEM images were taken with a Hitachi S-5500. Samples for STEM measurements were drop cast on copper TEM grids with carbon supports (400 mesh, TedPella). NCs were sized by image analysis (shown below). Samples for top-down and cross-section film SEM were prepared by spin coating ITO NC dispersions in a mixed solvent of 1:1 hexane:octane on undoped silicon substrates and processed as discussed in main text.

### Inductively coupled plasma-atomic emission spectroscopy (ICP-AES) measurement

The overall tin dopant concentration of ITO NCs were characterized by ICP-AES on a Varian 720-ES ICP Optical Emission Spectrometer after digesting the NCs with nitric acid.

### X-ray diffraction (XRD) measurement

XRD patterns of ITO NC films were collected on a Rigaku Miniflex 600 diffractometer using Cu K$_\alpha$ radiation. Films were identical to those imaged by SEM. NCs were sized using the Debye-Scherrer equation, $t = \frac{\lambda}{\beta \cos(\theta)}$, where $t$ is the NC diameter, $\lambda$ is the x-ray wavelength (0.15418 nm), $\beta$ is the full-width at half-max (FWHM) of the XRD peak, and $\theta$ is the Bragg angle of the XRD peak. The FWHM is corrected for instrumental broadening by, $\beta = \left(w_{\exp}^2 - w_{ins}^2\right)^{\frac{1}{2}}$, where $w_{exp}$ is the experimental XRD peak FWHM and $w_{ins}$ is the instrumental broadening as measured from the FWHM of LaB$_6$.

### X-ray photoelectron spectroscopy measurement

XPS was conducted on Kratos x-ray photoelectron spectrometer – axis ultra DLD using Al K$_\alpha$ x-ray source. Samples were identical to those imaged by SEM. XPS spectra were calibrated to the C 1s peak at 284.8 eV. Analysis was done using CasaXPS software. Near surface dopant concentration was assessed by calculating the ration of Sn 3d peak area to the total metal peak area.

### Visible spectrum spectroscopy measurement

Visible transparency was measured conducted on an Agilent Cary series UV-vis-NIR spectrophotometer. Samples used for visible range spectroscopy were the same samples in quartz that were used for conductivity measurements.

### Ellipsometric porosimetry measurement

Spectroscopic ellipsometry data was obtained using a J. A. Woollam M-2000 Spectroscopic Ellipsometer DI from 193 nm to 1690 nm. The J. A. Woollam Environment Cell was used to change the relative partial pressure of toluene while collecting ellipsometry data to monitor changes in film optical constants as the medium dielectric is modulated in a controlled manner. The angle of incident light was fixed at 70° relative to the sample, which corresponds to a normal incidence with respect to the cell window. Optical constants were obtained by fitting spectroscopic ellipsometry data over the spectral range of 400 nm to 1000 nm in CompleteEASE software and approximating the film as a Cauchy oscillator. The Lorentz-Lorentz equation was used to calculate the volume of toluene in film pores as a function of toluene partial pressure, which was fed into an effective medium approximation. Evolution of film optical constants with toluene partial pressure was fit for film porosity.

**Temperature dependent conductivity measurements**

Temperature dependent conductivity measurements were done on a Quantum Design Physical Properties Measurement System with external electronics sources. Current from -1 to 1 uA in 500 nA increments was applied through adjacent corners of the sample using a Keithley 220 current source and the voltage measured on the other two corners using a Keithley 2182 nanovoltmeter. A Keithley 2700 switch box was used to perform 4 point measurements in both configurations, and ohmic contact in 4 point and 2 point configurations was ensured at 300K and the lowest temperature measured. The conductance values for the two configurations were then used to compute the sheet conductance numerically using the Van der Pauw formula. The sheet conductivity data presented in the paper was collected while warming up, and temperature was held at each temperature while the IV sweep was performed. The R vs T curve was compared with the cool down curve to ensure no time dependence on R.

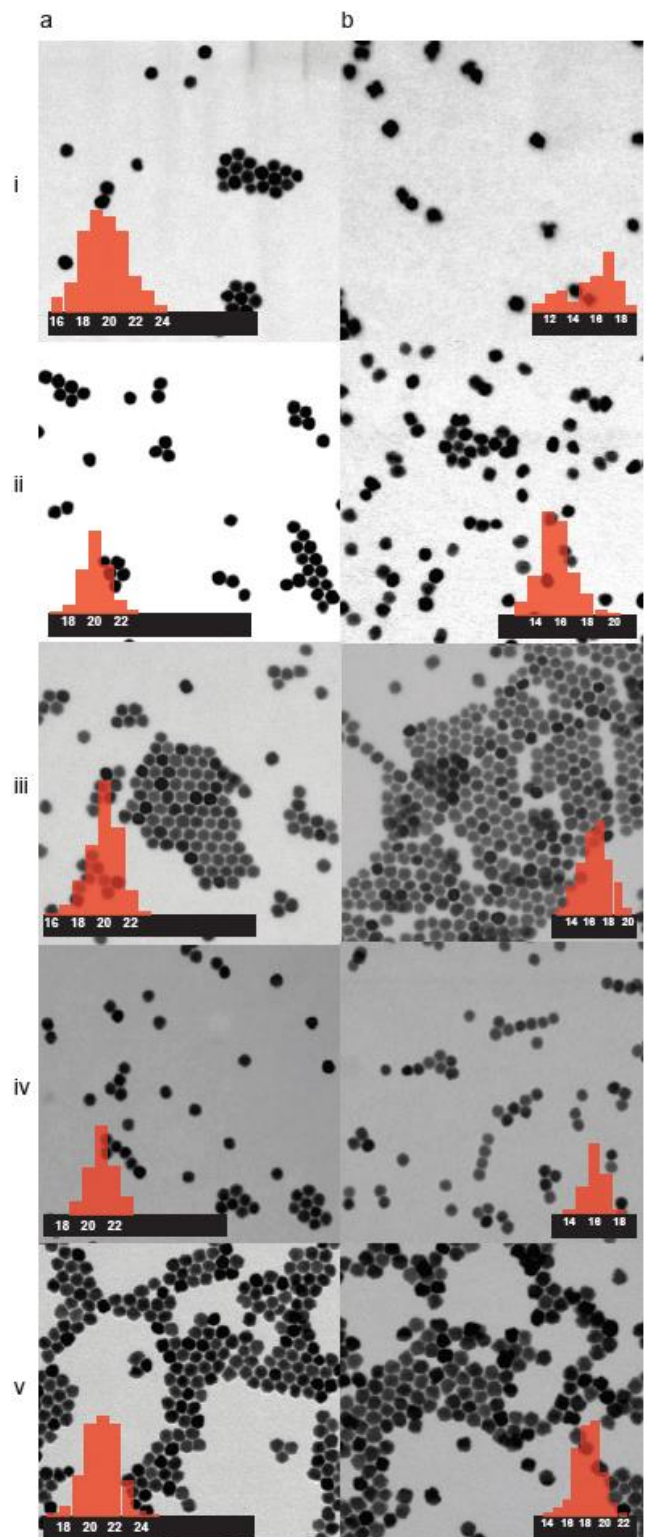

**Figure S1. Scanning transmission electron microscopy (STEM) images** of overall ITO NCs (a) and cores (b) for Core8 (i), Core5 (ii), Uniform (iii), Shell5 (iv), and Shell8 (v). Scale bars represent 100nm.

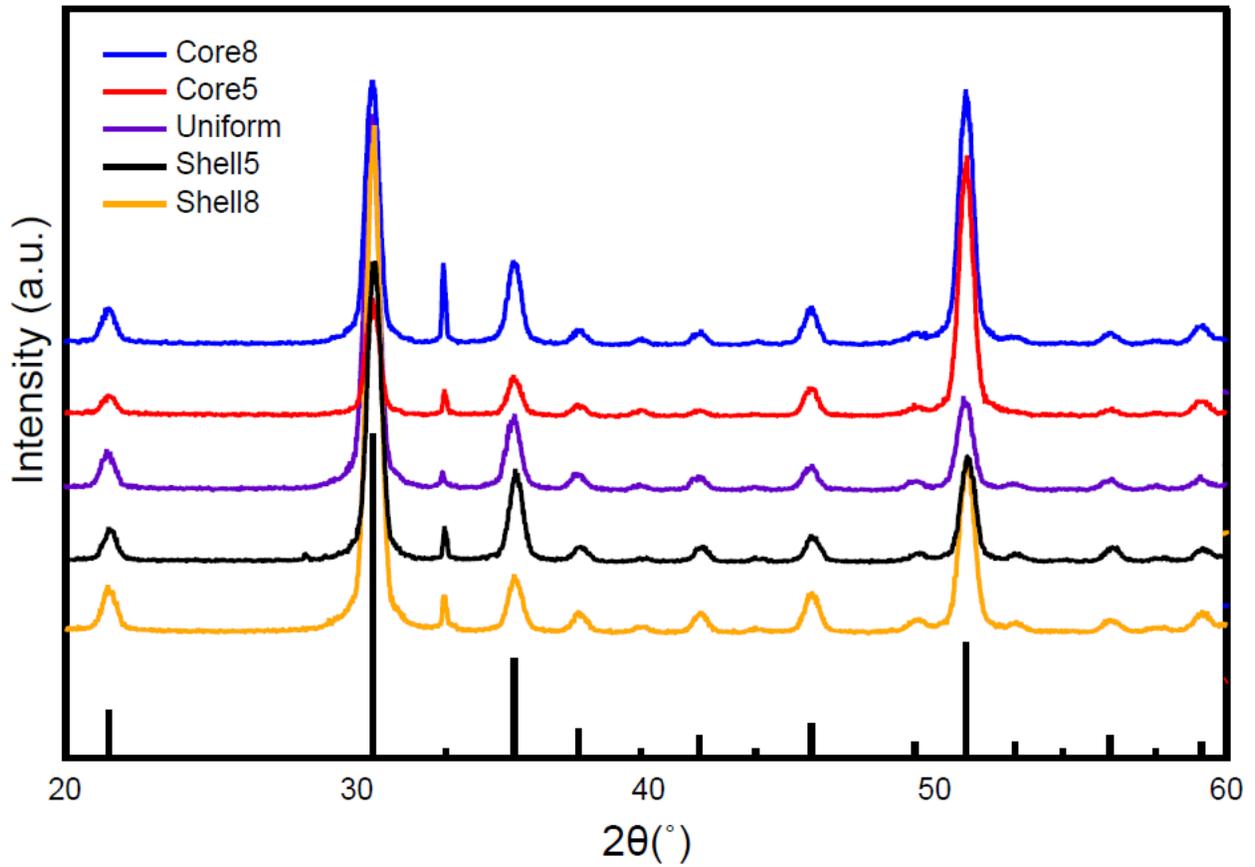

**Figure S2. Overall ITO NC crystal structure.** All samples show agreement with bixbyite ITO reference (sticks shown above along x-axis).

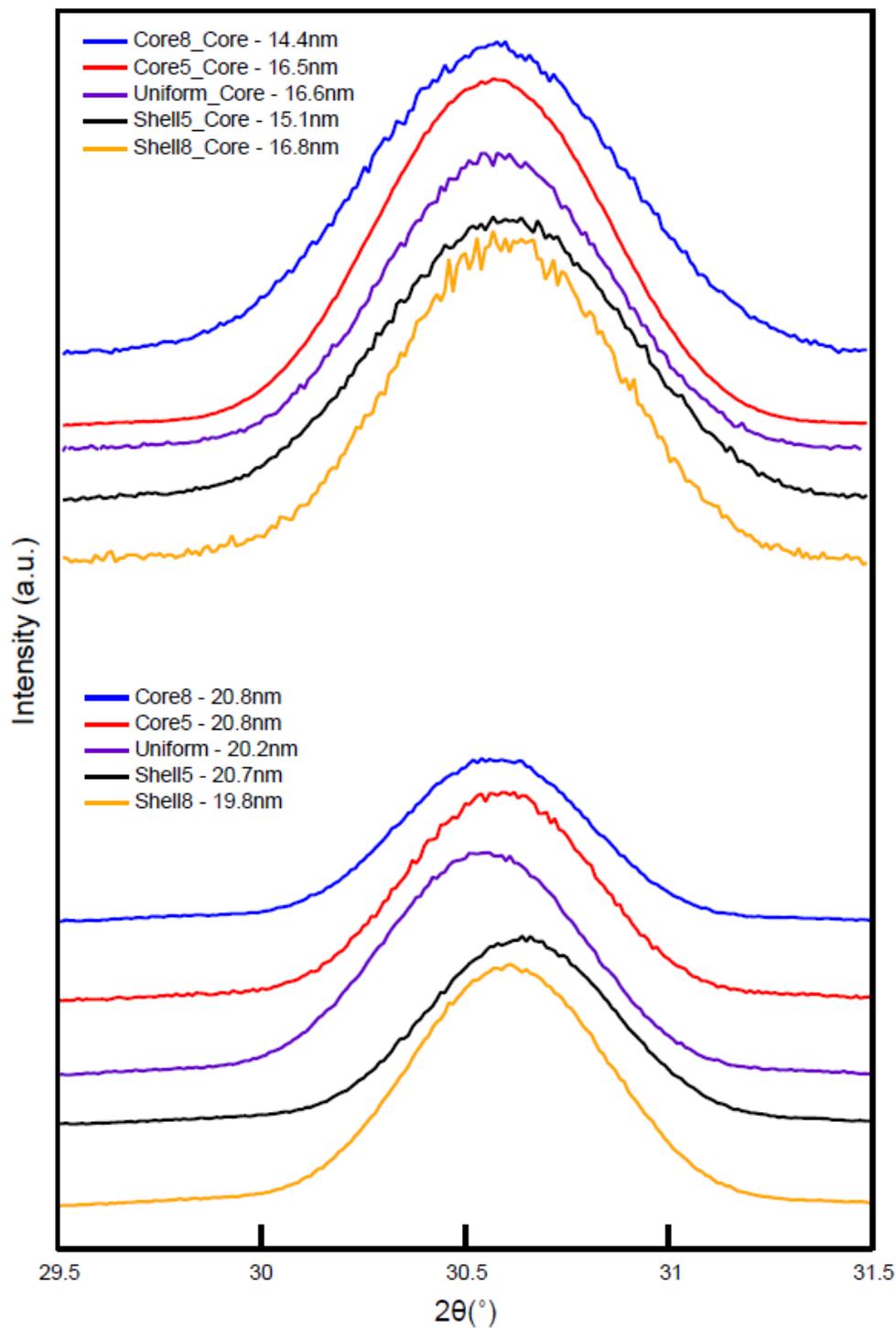

**Figure S3. (222) XRD peaks** of ITO NC cores (a) and overall ITO NCs (b). NC sizes were assessed by Scherrer analysis.

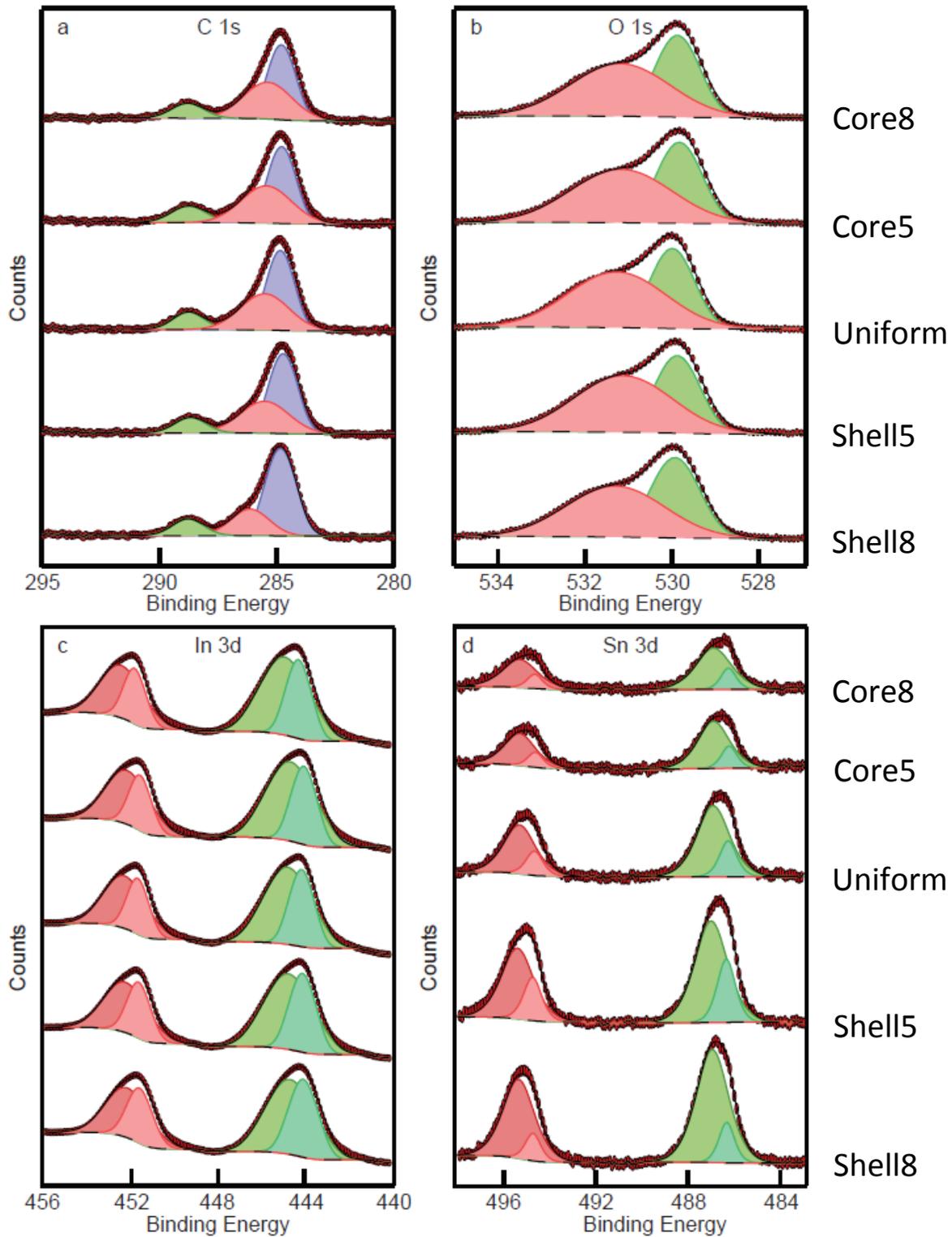

**Figure S4. X-ray photoelectron spectroscopy (XPS)** of C 1s (a), O 1s (b), In 3d (c), and Sn 3d (d) peaks for ITO NC films.

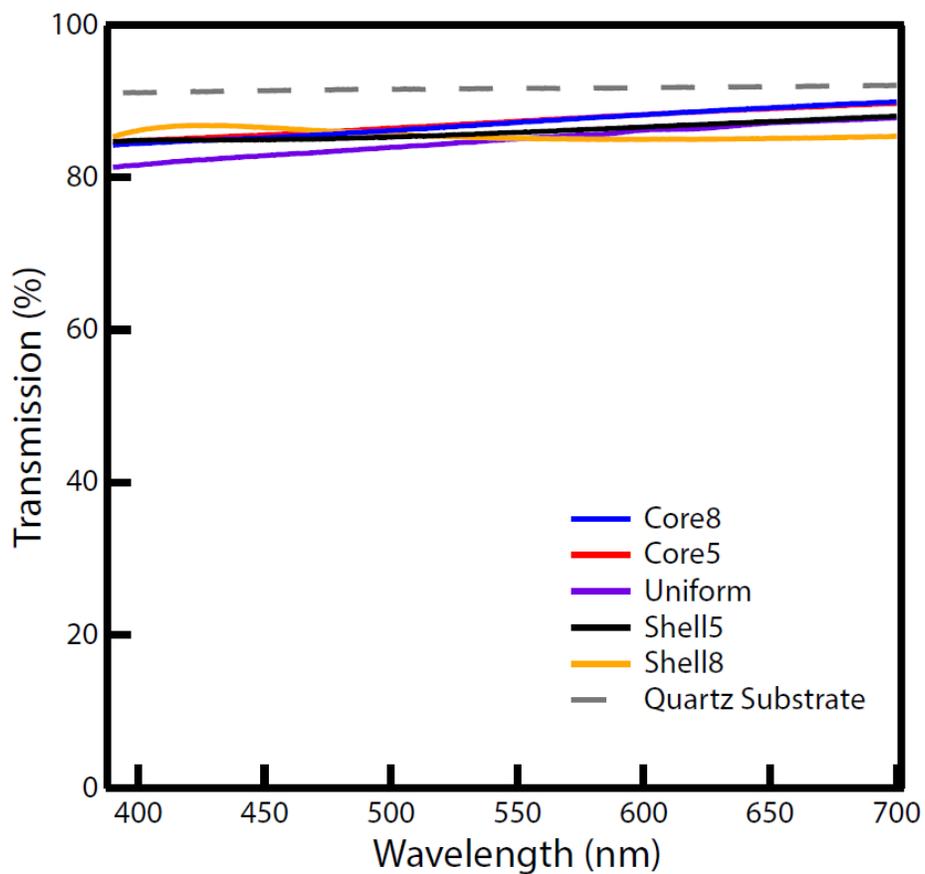

**Figure S5. Bare ITO NC film optical transparency.** All measured films show a wide transparency window across the visible range.

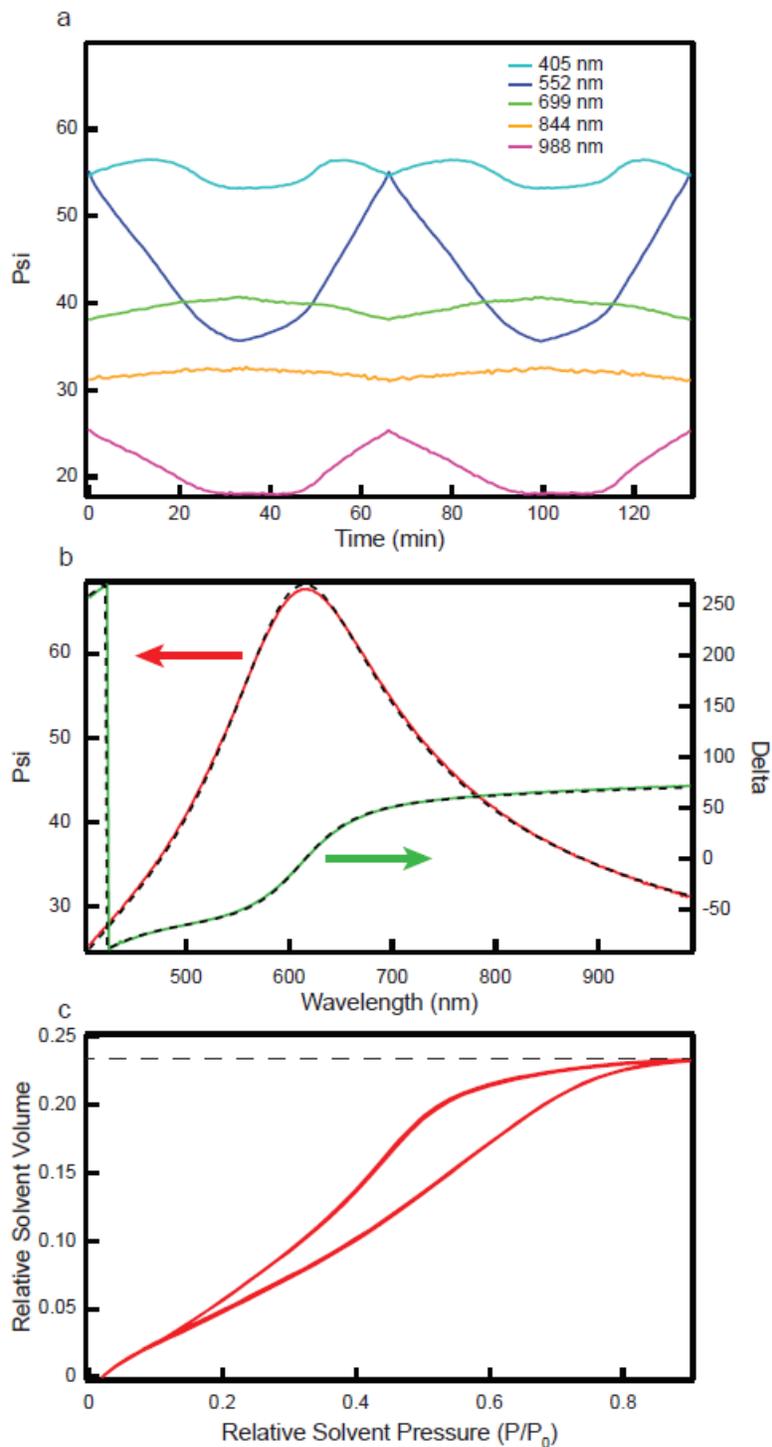

**Figure S6. Ellipsometric porosimetry (EP) data and analysis of Shell8.** EP data was collected from 0.02 to 0.9 relative solvent pressure with toluene as the solvent. Samples were cycled twice to ensure samples were not altered as a result of solvent exposure. Evolution of Psi with time (relative solvent pressure) (a), initial Psi and Delta with fits (dashed line) (b). Relative solvent volume as a function of solvent pressure (c). The maximum relative solvent volume is the sample porosity as shown by the dashed line.

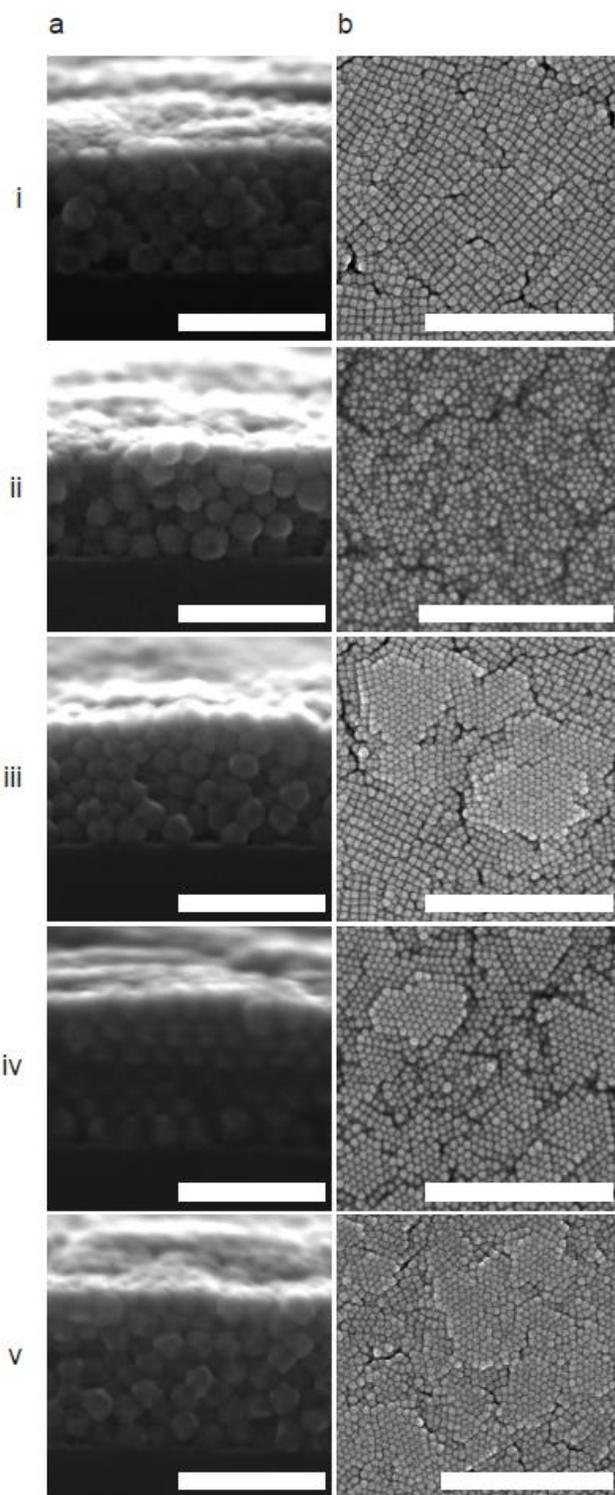

**Figure S7. Alumina-capped NC film on silicon SEM** cross-section (a) and top-down (b) for Core8 (i), Core5 (ii), Uniform (iii), Shell5 (iv), and Shell8 (v). Scale bars represent 100nm for cross-section and 500nm for top-down.

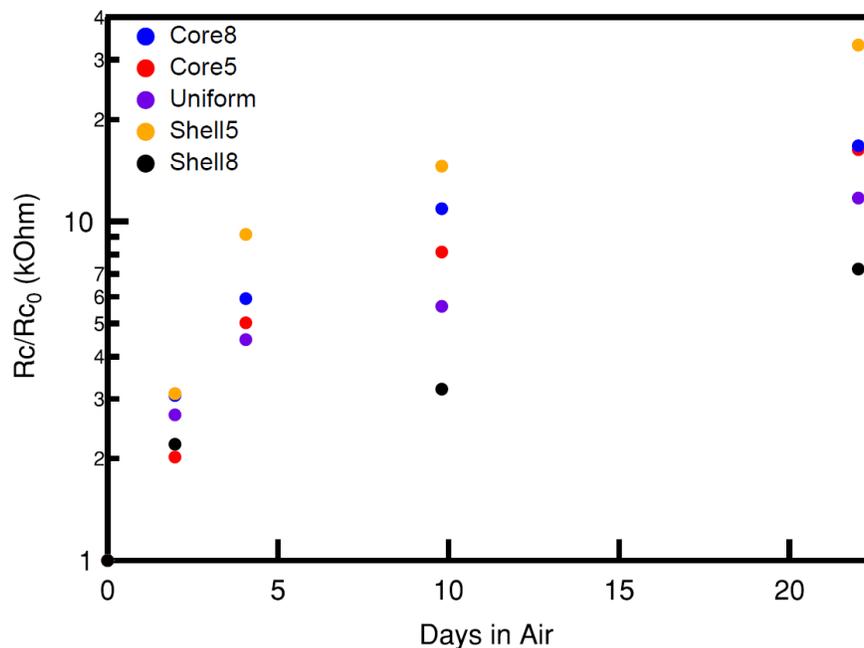

**Figure S8. Bare ITO NC air sensitivity.** The resistivity of bare ITO NC films is highly sensitive to ambient air exposure. The air sensitivity of metal oxide NC film resistivity is attributed to adsorbed water species at the NC surface.[3,4] This can cause otherwise identical films to display significantly different resistivity data due to different air exposure times.

## SI Text 2. Critical tunneling resistance of bare ITO NCs.

The critical tunneling resistance defines the maximum tunneling resistance for a granular film to exhibit metallic conduction (or granular metal conduction). If the tunneling resistance between two NCs is greater than the critical tunneling resistance, the film is expected to behave as an insulator and conduction will proceed through a hopping mechanism. The critical tunneling resistance in units of $e^2/\hbar$ is defined by

$$R_{bond}^C = \frac{6\pi}{\ln\left(\frac{E_C}{\delta}\right)} \quad \text{S1}$$

Charging energy, $E_C = \frac{e^2}{2\pi\varepsilon_c\varepsilon_0\alpha}$ \quad S2

Where $e$ is the electron charge, $\varepsilon_0$ is the permittivity of vacuum, and $\alpha$ is the grain diameter

Effective dielectric constant, $\varepsilon_c = \frac{\pi}{2}\varepsilon_m\left(\frac{2\epsilon}{\pi\varepsilon_m}\right)^{\frac{2}{5}} = 1.99$ \quad S3

Where $\varepsilon_m$ is the dielectric constant of the medium (1 for air) and $\epsilon$ is the dielectric constant of the NC (9)

Mean energy spacing in a single grain, $\delta = \left(g_{E_F} V_{NC}\right)^{-1}$ \quad S4

Where $V_{NC}$ is the NC volume

The density of states at the Fermi level, $g_{E_F} = \frac{3^{\frac{1}{3}} m_e^* n_e^{\frac{1}{3}}}{\hbar^2 \pi^{\frac{4}{3}}}$ \quad S5

Where $m_e^*$ is the effective mass of an electron ($0.4 m_e$ for ITO), $\hbar$ is Planck's constant, and $n_e$ is the electron concentration.

We assume complete dopant activation such that electron concentration is defined by (3E20)*(at% Sn overall) where 3E20 is the density of indium atoms in ITO. Values from these calculations and experimental bond resistance are shown in Table SI 1.

**Table SI 1. Critical bond resistance.** $\alpha$ is the grain (NC) diameter, $n_e$ is the electron concentration, $g_{E_F}$ is the density of states at the Fermi level, $\delta$ is the mean energy spacing, $E_C$ is the charging energy, $R_{bond}^C$ is the critical bond resistance, $R_{bond}^{Exp}$ are calculated bond resistances from experimental values.

| Sample | $\alpha$ (nm) | $n_e$ (cm$^{-3}$) | $g_{E_F}$ (J$^{-1}$m$^{-3}$) | $\delta$ (J) | $E_C$ (J) | $R_{bond}^C$ (kΩ) | $R_{bond}^{Exp}$ (kΩ) |
|---|---|---|---|---|---|---|---|
| Core8 | 20.8 | 9.9E20 | 1.0228E46 | 2.08E-23 | 1.11E-20 | 38.8 | 1240 |
| Core5 | 20.8 | 9.3E20 | 1.00171E46 | 2.12E-23 | 1.11E-20 | 38.9 | 915 |
| Uniform | 19.4 | 9.0E20 | 9.9082E45 | 2.64E-23 | 1.19E-20 | 39.9 | 504 |
| Shell5 | 10.8 | 9.3E20 | 1.00171E46 | 2.12E-22 | 1.11E-20 | 38.9 | 152 |
| Shell8 | 19.8 | 7.5E20 | 9.32398E45 | 2.64E-23 | 1.17E-20 | 40.0 | 974 |

Table SI 1 indicates all samples measured are consistent with insulating behavior and expected to conduct electrons through a hopping mechanism.

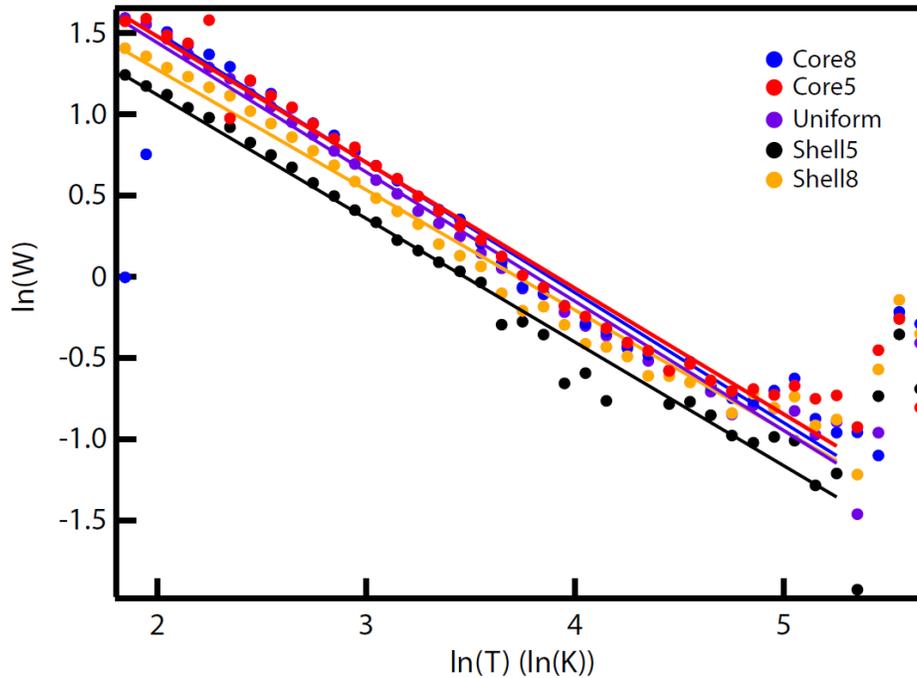

**Figure S9. Zabrodskii analysis** shows slopes near 0.78 for all samples measured. Exact values found are 0.78 (Core8), 0.78 (Core5), 0.78 (Uniform), 0.75 (Shell5), 0.70 (Shell8). This indicates a temperature dependence of $\sigma \propto \exp(-T^{-0.78})$.[5]

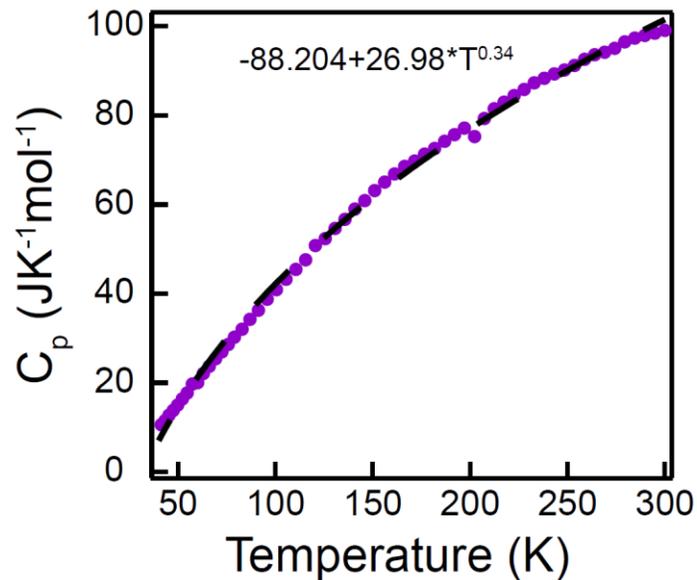

**Figure S10. ITO heat capacity.** Markers indicate data taken from Ref 6 and dashed line shows a power-law fit from ~40K to ~300K.[6]

**SI Text 3. Derivation of Efros-Shklovskii variable-range hopping with Gaussian broadened energy levels.**

Houtepen et. al. found the broadening of each energy level within the density-of-states of ensemble of nanocrystals due to temporal fluctuations was described by a Gaussian with width depending on temperature (T) and heat capacity ($C_V$) (S6).[7]

$$g(E) = g_0 \exp\left(-\frac{(E-E_0)^2}{2k_B T^2 C_V(T)}\right) \qquad \text{S6}$$

A materials heat capacity is temperature dependent for temperatures below its Debye temperature, $\theta_D$, (for ITO, $\theta_D$=1000K)[8]. If the heat capacity of approximated as a power-law in temperature where $Cp=C*T^m$, the tunneling rate is given by S7.

$$\Gamma \propto \exp\left(-\frac{2R}{a} - \frac{\Delta E^2 (1+m)}{4k_B T^{m+2} C}\right) \qquad \text{S7}$$

Where ΔE is the energetic barrier to hopping, a is the electron localization length, and R is the hopping distance, given by S8.

$$R = \frac{A}{\Delta E^n} \qquad \text{S8}$$

Where A and n are constants that depend on the hopping mechanism. By plugging S8 into S7 and maximizing Γ with respect to ΔE, we find the most optimal energetic barrier and hopping distance as a function of temperature (S9 & S10).

$$\Delta E_{opt} = \left(\frac{4AnCk_B T^{2+m}}{a(1+m)}\right)^{\frac{1}{2+n}} \qquad \text{S9}$$

$$R_{opt} = A\left(\frac{a(1+m)}{4AnCk_B T^{2+m}}\right)^{\frac{n}{2+n}} \qquad \text{S10}$$

Plugging S9 & S10 into S7 yields the general expression for hopping in systems with a Gaussian broadening of energy levels (S11).

$$\Gamma \propto \exp\left(-\left(\frac{(1+m)^n A^2 (2+n)^{2+n}}{a^2 4^n k_B^n C^n n^n T^{mn+2n}}\right)^{\frac{1}{2+n}}\right) \qquad \text{S11}$$

For Efros-Shklovskii variable-range hopping $\left(n = 1 \ \& \ A = \frac{e^2}{4\pi\varepsilon\varepsilon_0}\right)$ and m = 0.34 (Figure S9), the expression simplifies to S12.

$$\sigma \propto \Gamma \propto \exp\left(-\left(\frac{T_0}{T}\right)^{0.78}\right) \ where \ T_0 = \left(\frac{3.15 e^4}{16\pi^2 \varepsilon^2 \varepsilon_0^2 k_B C a^2}\right)^{\frac{1}{3}} \qquad \text{S12}$$

## SI Text 4. Simulation methods.

Poisson's equation was solved numerically for spherical nanocrystals with a given radial dopant profile and surface potential, $E_S$, using a finite element method. The charge density at any point inside the nanocrystal is made up of mobile electrons and immobile ionized impurity centers. Here, we have shown the potentials used to solve the Poisson's equation (Figure S11). $E_F$ is the Fermi energy level, $E_{CB}$ is the conduction band minima, $E_{VB}$ is the valence band maxima, $E_I$ is the reference potential and center of the band gap.

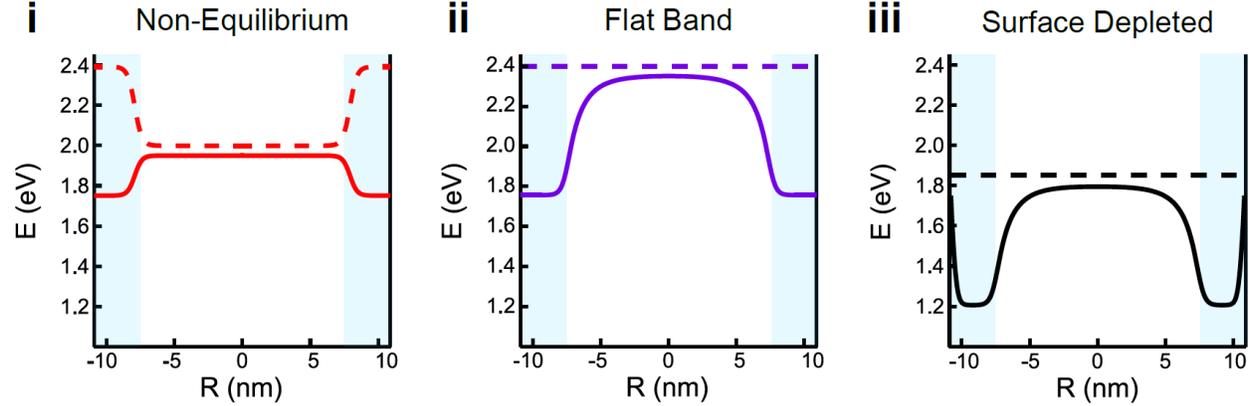

**Figure S11: Band profile of Core5** under non-equilibrium (i), flat band (ii), and surface depleted (iii) conditions. The non-equilibrium case shows the band profile and Fermi level before the equilibration of the Fermi level. The flat band case shows the core-shell equilibrium band profile with passivated surfaces, i.e. the surface potential is equal to the Fermi level of the shell species). The surface depleted case is the result of imposing a surface potential 0.2 eV below the flat band potential of indium oxide on case (ii).

Here, we adapted the dimensionless form of Poisson's equation derived by Seiwatz and Green[9] to solve numerically for a spherical nanoparticle in Cartesian coordinates as,

$$\nabla^2 u = -\frac{e^2 \rho}{\varepsilon \varepsilon_0 k_B T} \qquad \text{S13}$$

The non-dimensional potential is defined as, $u = \frac{E_F - E_I}{k_B T}$, and $k_B$ is the Boltzmann constant and $T$ is temperature. $\varepsilon_0$ is the vacuum permittivity, $\varepsilon$ is the static dielectric constant, and $\rho$ is the charge density.

$$\rho = \{\rho_D(r) - \rho_A(r) + p(r) - n(r)\} \qquad \text{S14}$$

where, $\rho_D(r)$ is the radially changing donor dopant density, $\rho_A(r)$ is the acceptor dopant density, $p(r)$ is hole density, $n(r)$ is electron density. Here, since we only have aliovalent donor dopants, $\rho_A(r) = 0$.

The free electron concentration in the parabolic conduction band is equal to,

$$n(r) = 4\pi \left(\frac{2m_e k_B T}{h^2}\right)^{\frac{3}{2}} \left(F_{\frac{1}{2}}\left(u - w_{C,I}(r)\right)\right) \qquad \text{S15}$$

where $F_{\frac{1}{2}}(\eta) = \int_0^\infty \frac{x^{\left(\frac{1}{2}\right)} dx}{1 + \exp(x - \eta)}$ and $w_{C,I} = \frac{E_{CB}(r) - E_I}{k_B T}$

Similarly, hole concentration in the parabolic valence band is equal to

$$p(r) = 4\pi \left(\frac{2m_h k_B T}{h^2}\right)^{\frac{3}{2}} \left(F_{\frac{1}{2}}\left(w_{V,I}(r) - u\right)\right) \qquad \text{S16}$$

where $w_{V,I} = \frac{E_{VB}(r) - E_I}{k_B T}$

If the donor energy level is $E_D$, the activated dopant concentration can be expressed as,

$$\rho_D(r) = \frac{N_D(r)}{1+2\exp(u-w_{D,I})} \qquad \text{S17}$$

where $w_{D,I} = \frac{E_D(r)-E_I}{k_B T}$

Substituting all the individual terms into Equation S13

$$\nabla^2 u = -\frac{e^2}{\varepsilon\varepsilon_0 k_B T}\left\{\frac{N_D(r)}{1+2\exp(u-w_{D,I})} + 4\pi\left(\frac{2m_h k_B T}{h^2}\right)^{\frac{3}{2}}\left(F_{\frac{1}{2}}(w_{V,I}(r)-u)\right) - 4\pi\left(\frac{2m_e k_B T}{h^2}\right)^{\frac{3}{2}}\left(F_{\frac{1}{2}}(u-w_{C,I}(r))\right)\right\} \qquad \text{S18}$$

with the boundary condition,

$$u = u_{surf} = \frac{E_{surf}-E_I}{k_B T} \qquad \text{S19}$$

Poisson's equation (Equation S18) was solved in COMSOL using a finite element scheme.

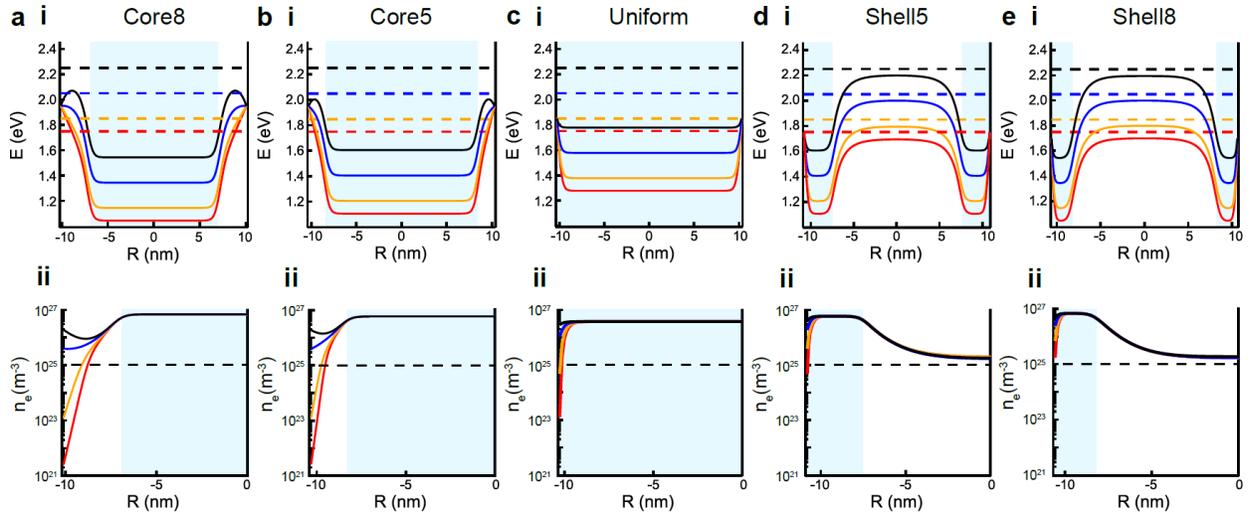

**SI Figure12. Band profiles under various surface potentials.** Intra-NC band (i) and radial electron concentration (ii) profiles for Core8 (a), Core5 (b), Uniform (c), Shell5 (d), and Shell8 (e) with a surface potential 0.3 eV below (red), 0.2 eV below (orange), 0 eV above (blue), 0.2 eV above (black) the flat band potential of indium oxide. The horizontal dashed line is the Fermi level (i) or critical carrier concentration (ii). In all cases, the blue shaded region indicates the region of enriched dopants. R = 0 denotes the center of a NC and maximum |R| denotes the surface. For all simulations showing a surface state below the flat band potential of indium oxide the absolute surface state energy does not change the qualitative interpretation of the data. This range of surface state energy is consistent with literature values of surface hydroxyls, ranging from 0.1 to 1 eV below the indium oxide conduction band minimum.[3,10]

## SI Text 5. Granular metal model validity in alumina-capped ITO NC Films.

There are three criteria for the granular metal conduction model to be valid:[11,12]
1. The intra-grain conductance is greater than the tunneling conductance ($g_0 \gg g_T$)
2. The tunneling conductance is greater than the critical tunneling conductance ($g_T > g_T^C$)
3. Must be in the high temperature limit to ignore quantum effects ($T > g_T \delta$)

Each of these criterion is addressed following equations presented by Beloborodov et. al.[12]

1. $g_0 > g_T$

The intra-grain conductance in units of $e^2/\hbar$ is defined by

$$g_0 = \frac{E_{Th}}{\delta}$$

Thouless energy, $E_{Th} = \frac{4\hbar D_0}{\alpha^2}$      S20

Classical diffusion coefficient, $D_0 = \frac{v_F^2 \tau}{d}$      S21

Where d is grain dimensionality (3 here)

Fermi velocity, $v_F = \frac{\hbar}{m_e^*}(3\pi^2 n_e)^{\frac{1}{3}}$

Taking the mean free path of an electron to be approximately the diameter of a single grain,

$$\tau = \frac{\alpha}{v_F}$$      S22

Plugging S15 into S14,

$$D_0 = \frac{v_F \alpha}{d}$$      S23

Values from these calculations and tunneling conductance are shown in Table SI 2.

**Table SI 2. Intra-NC and tunneling conductance.** $v_F$ is the Fermi velocity, $D_0$ is the classical diffusion coefficient for electrons, $E_{Th}$ is the Thouless energy, $g_0$ is the intra-NC conductance, and $g_T$ is the tunneling conductance from the granular metal fit.

| Sample | $v_F$ (m/s) | $D_0$ (m²/s) | $E_{Th}$ (J) | $g_0$ ($e^2/\hbar$) | $g_T$ ($e^2/\hbar$) |
|---|---|---|---|---|---|
| Core8 | 8.93E5 | 6.19E-3 | 3.81E-20 | 1840 | 0.641 |
| Core5 | 8.74E5 | 6.06E-3 | 3.73E-20 | 1760 | 0.675 |
| Uniform | 8.65E5 | 5.59E-3 | 3.96 E-20 | 1500 | 0.623 |
| Shell5 | 8.74E5 | 6.06E-3 | 3.73E-20 | 1760 | 0.654 |
| Shell8 | 8.14E5 | 5.37E-3 | 3.65E-20 | 1380 | 0.538 |

Tunneling conductance is much lower than intra-grain conductance for all measured samples. This criteria shows that tunneling between grains in the rate limiting step in electron conduction through these films.

2. ($g_T > g_T^C$)

The critical tunneling conductance in units of $e^2/\hbar$ is defined by

$$g_T^C = \frac{\ln\left(\frac{E_C}{\delta}\right)}{6\pi} = \frac{1}{R_{bond}^C}$$

Using equations equations S1 – S5 with $\varepsilon_m = 9$ for alumina, $g_T^C$ and $g_T$ are shown in Table SI 3.

**Table SI 3. Critical and experimental tunneling conductance.** $g_T^C$ is the critical tunneling conductance and $g_T$ is the tunneling conductance from the granular metal fit.

| Sample | $g_T^C$ ($e^2/\hbar$) | $g_T$ ($e^2/\hbar$) |
|---|---|---|
| Core8 | 0.265 | 0.641 |
| Core5 | 0.264 | 0.675 |
| Uniform | 0.256 | 0.623 |
| Shell5 | 0.264 | 0.654 |
| Shell8 | 0.255 | 0.538 |

All samples measured show tunneling conductance well above the critical tunneling conductance. The critical tunneling conductance defines the criterion for a material to behave as a metal.

3. ($T > g_T \delta$)

Using equations S4 and S5 as well as experimental $g_T$, $g_T \delta$ is shown in Table SI 4.

**Table SI 4. Quantum temperature upper limit.**

| Sample | $g_T \delta$ (K) |
|---|---|
| Core8 | 0.96 |
| Core5 | 1.04 |
| Uniform | 1.19 |
| Shell5 | 1.00 |
| Shell8 | 1.03 |

Temperature $g_T \delta$ defines the temperature above which quantum effects no longer significantly influence electron physics in a material. All fits in this work were done at or above 2K. Therefore, we do not expect significant quantum effects.

**References:**

(1) Jansons, A. W.; Hutchison, J. E. Continuous Growth of Metal Oxide Nanocrystals: Enhanced Control of Nanocrystal Size and Radial Dopant Distribution. *ACS Nano* **2016**, *10*, 6942–6951.

(2) Crockett, B. M.; Jansons, A. W.; Koskela, K. M.; Johnson, D. W.; Hutchison, J. E. Radial Dopant Placement for Tuning Plasmonic Properties in Metal Oxide Nanocrystals. *ACS Nano* **2017**, *11*, 7719–7728.

(3) Thimsen, E.; Johnson, M.; Zhang, X.; Wagner, A. J.; Mkhoyan, K. A.; Kortshagen, U. R.; Aydil, E. S. High Electron Mobility in Thin Films Formed via Supersonic Impact Deposition of Nanocrystals Synthesized in Nonthermal Plasmas. *Nat. Commun.* **2014**, *5*.

(4) Ephraim, J.; Lanigan, D.; Staller, C.; Milliron, D. J.; Thimsen, E. Transparent Conductive Oxide Nanocrystals Coated with Insulators by Atomic Layer Deposition. *Chem. Mater.* **2016**, *28*, 5549–5553.

(5) Zabrodskii, A. G. The Coulomb Gap: The View of an Experimenter. *Philos. Mag. Part B* **2001**, *81*, 1131–1151.

(6) Cordfunke, E. H. P.; Westrum, Jr, E. F. The Heat Capacity and Derived Thermophysical Properties of In2O3 from 0 to 1000 K. *J. Phys. Chem. Solids* **1992**, *53*, 361–365.

(7) Houtepen, A. J.; Kockmann, D.; Vanmaekelbergh, D. Reappraisal of Variable-Range Hopping in Quantum-Dot Solids. *Nano Lett.* **2008**, *8*, 3516–3520.

(8) Lin, J.-J.; Li, Z.-Q. Electronic Conduction Properties of Indium Tin Oxide: Single-Particle and Many-Body Transport. *J. Phys. Condens. Matter* **2014**, *26*.

(9) Space Charge Calculations for Semiconductors. *J. Appl. Phys.* **1958**, *29*, 1034–1040.

(10) Zandi, O.; Agrawal, A.; Shearer, A. B.; Gilbert, L. C.; Dahlman, C. J.; Staller, C. M.; Milliron, D. J. Impacts of Surface Depletion on the Plasmonic Properties of Doped Semiconductor Nanocrystals. *ArXiv170907136 Cond-Mat Physicsphysics* **2017**.

(11) Beloborodov, I. S.; Efetov, K. B.; Lopatin, A. V.; Vinokur, V. M. Transport Properties of Granular Metals at Low Temperatures. *Phys. Rev. Lett.* **2003**, *91*, 246801.

(12) Beloborodov, I. S.; Lopatin, A. V.; Vinokur, V. M.; Efetov, K. B. Granular Electronic Systems. *Rev. Mod. Phys.* **2007**, *79*, 469–518.